\documentclass[11pt]{amsart}
\usepackage{amsbsy,amssymb,amscd,amsfonts,latexsym,amstext,delarray,
amsmath,graphicx} 
\input xypic

\usepackage{tikz-cd}

\usepackage{color}

\newtheorem{thm}{Theorem}[section]
\newtheorem{prop}[thm]{Proposition}
\newtheorem{cor}[thm]{Corollary}
\newtheorem{lem}[thm]{Lemma}

\newtheorem{defn}[thm]{Definition}
\newtheorem{rem}[thm]{Remark}
\newtheorem{ex}[thm]{Example}

\numberwithin{equation}{section}

\def\bH{{\mathbb H}}

\def\C{{\mathbb C}}
\def\F{{\mathbb F}}
\renewcommand{\H}{{\mathbb H}}
\def\N{{\mathbb N}}

\def\Q{{\mathbb Q}}
\def\R{{\mathbb R}}
\def\Z{{\mathbb Z}}

\def\cA{{\mathcal A}}
\def\cB{{\mathcal B}}

\def\cF{{\mathcal F}}
\def\cG{{\mathcal G}}
\def\cH{{\mathcal H}}
\def\cI{{\mathcal I}}

\def\cK{{\mathcal K}}

\def\cN{{\mathcal N}}

\def\cP{{\mathcal P}}

\def\cR{{\mathcal R}}
\def\cS{{\mathcal S}}
\def\cT{{\mathcal T}}

\def\cV{{\mathcal V}}
\def\cW{{\mathcal W}}
\def\cX{{\mathcal X}}

\def\Aut{{\rm Aut}}

\def\GL{{\rm GL}}
 
\def\Hom{{\rm Hom}}

\def\Ker{{\rm Ker}}

\def\SL{{\rm SL}}
\def\Spec{{\rm Spec}}

\def\Tr{{\rm Tr}}

\def\fb{{\mathfrak b}}

\def\mmu{{\mathfrak r}}

\title{Quantum statistical mechanics in arithmetic topology}
\author{Matilde Marcolli and Yujie Xu}
\address{Division of Physics, Mathematics, and Astronomy, California Institute of Technology, 1200 E California Blvd, Pasadena, CA 91125, USA}
\email{matilde@caltech.edu}
\email{yujiex@caltech.edu}

\begin{document}
\maketitle

\begin{abstract}
This paper provides a construction of a quantum statistical mechanical system
associated to knots in the $3$-sphere and cyclic branched coverings of the $3$-sphere,
which is an analog, in the sense of arithmetic topology, of the Bost--Connes system,
with knots replacing primes, and cyclic branched coverings of the $3$-sphere
replacing abelian extensions of the field of rational numbers. The operator algebraic
properties of this system differ significantly from the Bost--Connes case, due to
the properties of the action of the semigroup of knots on a direct
limit of knot groups.  The resulting algebra of observables is a noncommutative
Bernoulli product. We describe the main properties of the associated
quantum statistical mechanical system and of the relevant partition functions,
which are obtained from simple knot invariants like genus and crossing number.
\end{abstract}

\section{Introduction}

This paper addresses a question asked to the first author by Masanori Morishita,
on the possibility of adapting to $3$-manifolds the Bost--Connes construction 
\cite{BC} of a quantum statistical mechanical system associated to the abelian 
extensions of $\Q$, and its generalizations to number fields \cite{CMR}, \cite{CorMa},
\cite{HaPa}, \cite{LLN}, along the lines of the general ``arithmetic topology" program.
The latter can be seen as a broad dictionary of analogies between the geometry
of knots and $3$-manifolds and the arithmetic of number fields, with knots as analogs
of primes and $3$-manifolds, seen as branched coverings of the $3$-sphere, viewed
as analogs of number fields. In this paper we answer Morishita's question by providing explicit
constructions of quantum statistical mechanical systems associated to (alternating)
knots, to knot groups, and to cyclic branched covers of the $3$-sphere, with the
latter providing our analog of the abelian extensions of $\Q$ in the Bost--Connes
construction. The structure of the resulting quantum statistical mechanical 
systems is different from the Bost--Connes case and it leads to an algebra of
observables that can be expressed in the form of a Bernoulli crossed product,
of the type studied in noncommutative Bernoulli actions in the theory of factors. 
We relate the geometry and dynamics of our system to known invariants of knots 
and $3$-manifolds. 

\subsection{The principle of Arithmetic Topology}

Arithmetic topology originates from insights by John Tate and Michael Artin on
topological interpretations of class field theory. The analogy between primes
and knots, which is the founding principle of Arithmetic Topology, was first
observed by Barry Mazur, David Mumford, and Yuri Manin. The subject developed over the years,
with various contributions, such as \cite{Den}, \cite{Kap}, \cite{Kim}, \cite{Mori}, \cite{Mori2}, 
\cite{Rama}, \cite{Rez}, \cite{Sik}, as a powerful guiding principle
outlining parallel results and analogies between the arithmetic of number fields and
the topology of $3$-manifolds. The basic analogy sees number fields as analogs
of compact oriented $3$-manifolds, with $\Q$ playing the role of the $3$-sphere $S^3$.
Here the main idea is that, while number fields are finite extensions of $\Q$, ramified 
at a finite set of primes, all compact oriented $3$-manifolds can be described as
branched coverings of the $3$-sphere, branched along a link. A major point where this 
analogy does not carry over is the fact that, while the description of a number field
as ramified covering of $\Q$ is unique, there are many inequivalent ways of
describing $3$-manifolds as branched covers of the $3$-sphere, branched along
knots or links (or more generally embedded graph). While this lack of uniqueness
for $3$-manifolds can be used to make the construction dynamical, see \cite{MarZa},
the same dynamics does not apply to number fields. However, the corresponding
analogy between knots and primes, that results from this first analogy between
number fields and $3$-manifolds, has been very fruitful, leading to many new results,
ranging from arithmetic analogs for higher linking numbers \cite{Mori}, \cite{Mori2}, to
arithmetic Chern--Simons theory, \cite{Kim}. 

\smallskip

Over the past two decades, the connection between number theory and
quantum statistical mechanics was also widely explored, starting with
early constructions of statistical systems associated to the primes, 
\cite{Julia}, \cite{Spector}, the more refined Bost--Connes system \cite{BC}
which also involves the Galois theory of abelian extensions of $\Q$,
and subsequent generalizations of this construction to arbutrary
number fields, obtained in \cite{HaPa} and further studied in 
\cite{CorMa}, \cite{CorLiMa}, \cite{LLN}, \cite{NeRu}. The purpose of the present paper
is to recast the Bost--Connes construction in the setting of arithmetic topology,
with the semigroup of knots with the connecting sum operation replacing the
multiplicative semigroup of positive integers, and the cyclic branched coverings
of the $3$-spheres replacing the abelian coverings of $\Q$.

\subsection{Bost--Connes system}

We recall briefly the construction of the Bost--Connes algebra 
and quantum statistical mechanical system from \cite{BC} (see also \cite{CoMa} and \S 3 of \cite{CM-book}).
Consider the group ring $\Q[\Q/\Z]$ with generators $e(r)$ with
$r\in \Q/\Z$. 
The maps $\{ \sigma_n \}_{n\in \N_\rho}$ given by
\begin{equation}\label{sigmanBC}
\sigma_n(e(r)):=e(nr) 
\end{equation}
determine an action of the semigroup $\N$ by endomorphisms 
of the group ring $\Q[\Q/\Z]$. These endomorphisms have partial inverses
$\alpha_n: \Q[\Q/\Z] \to \Q[\Q/\Z]$,
\begin{equation}\label{alphanhatpi}
\alpha_n (e(r)) = \frac{1}{n} \sum_{s:\, ns=r} e(s)
\end{equation}
with $\sigma_n \circ \alpha_n (e(r)) =e(r)$
and $\alpha_n \circ \sigma_n (e(r)) = e_n \cdot e(r)$,
with $e_n = n^{-1} \sum_{s:\, ns=0} e(s)$ an idempotent in $\Q[\Q/\Z]$.
Thus, one can define the semigroup crossed product. This is the
(rational) Bost--Connes algebra $\cA_{BC,\Q}=\Q[\Q/\Z]\rtimes \N$
with generators $\mu_n$ and $e(r)$ and relations
\begin{equation}\label{relBC1}
 \mu_n^* \mu_n=1, \ \ \  \mu_n \mu_n^* =e_n, \ \ \ \mu_n \mu_m=\mu_{nm}, \ \ \ 
\mu_n \mu_m^* =\mu_m^* \mu_n \text{ for } (n,m)=1, 
\end{equation}
\begin{equation}\label{relBC2}
 \mu_n e(r) \mu_n^* = \alpha_n (e(r)), \ \ \ 
\mu_n^* e(r) \mu_n =\sigma_n(e(r)). 
\end{equation}
The complexification $\cA_{BC,\C}=\cA_{BC,\Q}\otimes_\Q \C$ has
a $C^*$-algebra completion given by the semigroup crossed
product $\cA_{BC}=C^*(\Q/\Z)\times \N$, with the same generators
and relations. The time evolution of the Bost--Connes system
is defined by $\sigma_t(\mu_n)=n^{it} \mu_n$ and $\sigma_t(e(r))=e(r)$.
The algebra $\cA_{BC}$ has representations on the Hilbert space
$\ell^2(\N)$, parameterized by the choice of an element $u\in \hat\Z^*$,
of the form
\begin{equation}\label{BCreps}
\pi_u(e(r)) \epsilon_m = u(r)^m\, \epsilon_m, \ \ \  \pi_u(\mu_n)\epsilon_m = \epsilon_{nm},
\end{equation}
where $u(r)$ is a root of unity in $\C$ determined by the embedding of
$\Q/\Z \hookrightarrow \C$ specified by the choice of $u\in \hat\Z^*$,
where we identify $\hat\Z=\Hom(\Q/\Z,\Q/\Z)$.  

\smallskip

Given a pair $(\cA,\sigma)$ of a $C^*$-algebra and a time evolution
$\sigma: \R \to \Aut(\cA)$, a KMS$_\beta$ state for $(\cA,\sigma)$
is a continuous linear functional $\varphi_\beta: \cA \to \C$ satisfying normalization
$\varphi_\beta(1)=1$ and positivity $\varphi_\beta(a^* a)\geq 0$ (that is, a {\em state} 
on $\cA$) such that, for all $a,b\in \cA$ there is a function $F_{a,b}(z)$ that
is holomorphic on the strip $\cI_\beta=\{ z\in \C\,:\, 0< \Im(z)< \beta \}$ and
continuous on the boundary $\partial\cI_\beta$ of the strip, such that
\begin{equation}\label{KMSstate}
 F_{a,b}(t)=\varphi_\beta(a\sigma_t(b)), \ \ \ \ F_{a,b}(t+i\beta)=\varphi(\sigma_t(b) a).
\end{equation}
In other words, the failure of a KMS$_\beta$ to be a trace is measured by interpolation
by a holomorphic function. 

\smallskip

The KMS states of the Bost--Connes system
$(\cA_{BC},\sigma)$ are completely classified and given by the following list of cases (see \cite{BC}):
\begin{itemize}
\item for every $0< \beta \leq 1$ there is a unique KMS$_\beta$ state $\varphi_\beta$
determined by 
$$ \varphi_\beta (e(\frac{a}{b})) = \frac{f_{-\beta+1}(b)}{f_1(b)} $$
where $f_k(b)=\sum_{d|b} \mu(d) (b/d)^k$, with $\mu$ the M\"obius function;
\item for every $\beta > 1$, the extremal KMS$_\beta$ states are given by
Gibbs states determined by
\begin{equation}\label{BClow}
 \varphi_{\beta, u}(e(r))= \frac{{\rm Li}_\beta(u(r))}{\zeta(\beta)}, 
\end{equation} 
where ${\rm Li}_\beta$ is the polylogarithm function, $u(r)$ is a root of unity, for a given $u\in \hat\Z^*$,
and $\zeta(\beta)$ is the Riemann zeta function;
\item for $\beta=\infty$ the extremal KMS$_\beta$ states are determined by
$\varphi_{\infty,u}(e(r))=u(r)$.
\end{itemize}

\smallskip

The Bost--Connes system is related to the arithmetic of $\Q$ and the Galois theory
of its abelian extensions. Generalizations of this quantum statistical mechanical
system were constructed for arbitrary number fields in \cite{HaPa},
\cite{LLN}, \cite{NeRu}, and further studied in \cite{CorMa}, \cite{CorLiMa}. 

\subsection{Structure of the paper}

In \S \ref{QSMknotsSec} we develop a quantum statistical mechanics of knots.
There is a natural semigroup structure on knots. It is given by the operation of
connected sum defined on equivalence classes of {\em oriented knots}. 
This operation gives rise to an abelian semigroup $(\cK,\#)$, which is infinitely generated, with
generators the prime knots. Each knot has a unique prime decomposition
$K=K_1\# \cdots \# K_m$ for some $m$, with $K_j$ prime knots. We focus on
knot invariants that behave well with respect to the connected sum operation.
In particular, we focus on simple invariants such as the genus and the
crossing number. In the latter case, it is at present an open conjecture whether
the invariant is additive over connected sums for all knots, but the result is known
to hold for alternating knots. Therefore, in the paper we often restrict our
attention to alternating knots, purely for the purposes of using these results
about the crossing number. Conditionally to the above mentioned conjecture,
one can reformulate them in terms of the larger semigroup of knots. We construct
a Hamiltonian based on genus and crossing number and we estimate in
Theorem~\ref{ZetaCrg} the range of 
convergence of the partition function using results of \cite{StoTcherVdo},
\cite{StoVdo} on the rate of growth of multiplicities. We show the uniqueness of
KMS states for this system of knots without interaction in Proposition~\ref{KMSKa}
and we discuss the type III nature of the high temperature state in
Proposition~\ref{KMSKa} and Theorem~\ref{IIIq}. 

\smallskip

In \S \ref{interSec} we return to the original system without
interaction of \cite{Julia} and \cite{BC}, with
prime numbers contributing independent oscillators (in the
form of Toeplitz operators) and we discuss how one can
try to extend it from the multiplicative semigroup $\N$ of
positive integer to the group $\Q^*_+$ of positive rational
numbers. We show that the Hamiltonian can be extended so that
the corresponding partition function is again expressible in
terms of the Riemann zeta function. We show in \S \ref{semiKgroupGsec}
that the same construction extends to the case of the Grothendieck
group of the semigroup of (alternating) knots with the connected
sum. Again, this result relies on estimates of \cite{StoTcherVdo},
\cite{StoVdo} on the number of alternating knots with fixed
genus and crossing number. However, at the level of the
algebra of observables of the system, this generalization of
the Hamiltonian requires an extension of the algebra by
the spectral projections of the Hamiltonian, in order to
remain invariant under the time evolution. This extension
has the effect of making the time evolution inner, which
is not desirable from the operator algebra perspective.
We bypass this problem by considering more general 
systems with interaction involving both knots and $3$-manifolds.

\smallskip

In \S \ref{3mfldSec}, we introduce cyclic branched
coverings of $S^3$, branched along a knot. We discuss
the behavior of knot groups under connected sums of knots, 
and we construct a directed system of knot groups
over the semigroup of knots ordered by ``divisibility" 
with respect to the connected sum operation. 
We interpret the resulting direct limit as the knot group
of a wild knot. We also consider a projective
limit, related to changing the order of the cyclic
branched cover. 

\smallskip

In \S \ref{3mfldQSMSec} we construct a more
refined system, which is more similar in nature to the
Bost--Connes system and which involves not only knots
but also the cyclic branched covers of $S^3$.
We begin by investigating the action of the
semigroup of knots with connected sum on the
group algebra of the direct limit of the system of
knot groups considered in the previous section.
We show that, unlike the Bost--Connes case,
the endomorphisms $\sigma_K$ are injective and
not surjective. The resulting crossed product
system is then more similar to the generalization 
of the Bost--Connes considered in \cite{Mar},
in relation to the Habiro ring. In particular, we show
that the resulting crossed product algebra is
in fact a noncommutative Bernoulli shift
$$ \bigotimes_{g\in \cG_\cK} C^*_r(\pi) \, \rtimes \cG_\cK, $$
where $\cG_\cK$ is the Grothendieck group of the
semigroup of knots $(\cK,\#)$ and $\pi =\varinjlim_K \pi_K$
is the direct limit of the system of knot groups. The
action of $\cG_\cK$ is the Bernoulli action that
permutes the terms in the crossed product
$\otimes_g C^*_r(\pi)$. We then include the
datum of the branched covers, in the form
of a group homomorphism $\rho: \pi \to \Q/\Z$.
We construct a projective limit of groups $\hat\pi_{K,n}$
and $\hat \pi_n$, which correspond to adding $n$-th 
roots of the generators of the knot group. This
construction is modelled on the construction of
roots of Tate motives in \cite{LoMa}. This 
construction allows us to replace the algebra $C^*_r(\pi)$,
which encodes the information about the knot groups, but
not about the coverings, with the more refined
$C^*_r(\hat\pi_\rho)\rtimes_\alpha \N_\rho$, where $\hat\pi_\rho$
is the projective limit of the system of the $\pi_n$ and $\N_\rho$
is a subsemigroup of $\N$, given by those integers that
are relatively prime to $n_\rho$, which is the order of
the root of unity that is the image under the morphism
$\rho$ of the generators of the group $\pi$. 
The semigroup action of $\N_\rho$ on $C^*_r(\hat\pi_\rho)$
is modeled on the Bost--Connes action, by viewing
$\hat\pi_\rho$ as a fibered product inside $\pi \times \Q/\Z$. 

\smallskip

We then construct time evolutions,
first on the algebra $C^*_r(\hat\pi_\rho)\rtimes_\alpha \N_\rho$,
induced by the Bost--Connes time evolution on $C^*(\Q/\Z)\rtimes \N$,
and then on the tensor product $\otimes_g \cB_g$, with $g\in\cG_\cK$
and $\cB_g=C^*_r(\hat\pi_\rho)\rtimes_\alpha \N_\rho$. In this tensor
product case, we take on each factor a version $\sigma_{t,g}$ of the Bost--Connes time evolution,
with the Hamiltonian $H_{BC}$ scaled by a factor $f(g)$, for a 
function $f: \cG_\cK \to \N$. In Proposition~\ref{ZetasigmaG}, we
identify a summability condition on the function $f(g)$ that
guarantees that the Hamiltonian has a well defined partition
function, which is convergent for $\beta >1$. Here the trace
of the operator $e^{-\beta H}$ in the partition function is a 
combination of the operator trace on $\ell^2(\N_\rho)$ and the
von Neumann trace on the group algebra of $\hat\pi_\rho$.
We also show how the KMS states of the Bost--Connes determine
KMS states for the system $(\otimes_g \cB_g, \otimes_g \sigma_{t,g})$.
In particular the low temperature states give rise to KMS states $\Psi_{\beta,f}$
for this system that are Gibbs states with respect to the 
partition function and the trace described in Proposition~\ref{ZetasigmaG}.
We then consider the crossed product $(\otimes_g \cB_g)\rtimes \cG_\cK$
and we show that the KMS states $\Psi_{\beta,f}$ transform, under the action
$\alpha_h$ of $h\in \cG_\cK$ as $\Psi_{\beta,f}\circ \alpha_h = \Psi_{\beta,\alpha_{h^{-1}}(f)}$.

\smallskip

Restricting to the subsemigroup $\cK_a$ of alternating knots, we show
that the same estimates of \cite{StoTcherVdo},
\cite{StoVdo} on the number of alternating knots with fixed
genus and crossing number that we used in \S \ref{semiKgroupGsec},
and the result of Theorem~\ref{ZetaCrg}, imply that a function satisfying
the desired convergence properties can be constructed using the
crossing number and the genus of knots.

\smallskip

\smallskip
\section{Quantum statistical mechanics of knots}\label{QSMknotsSec}

Consider the semigroup $(\cK, \#)$ or ambient isotopy classes of
oriented knots with the connected sum operation. The primary
decomposition of knots states that every $K\in \cK$ can be
decomposed into a direct sum of prime knots. There are infinitely
many prime knots, hence the semigroup $\cK$ is a countably
generated free abelian semigroup. A choice of an enumeration
of the prime knots gives a (non-canonical) semigroup isomorphism 
of $(\cK,\#)$ with $(\N,\cdot)$ by mapping prime knots to
the prime numbers. The identification is non-canonical as prime
knots, unlike prime numbers, have no natural ordering. However,
this identification suggests that the quantum statistical mechanics
of creation-annihilation operators constructed out of the
primary decomposition in $\N$ (see 
\cite{Julia}, \cite{Spector}, and \S 2 of \cite{BC})
can be directly adapted to the semigroup of knots.

\smallskip

Let $\cP_\cK$ denote the set of prime knots. As in the case of 
the semigroup $\N$, we can identify $\ell^2(\cK)$ with the bosonic
Fock space $\ell^2(\cK)=\oplus_{n=1}^\infty S^n \ell^2(\cP_\cK)$,
where $S^n\cH$ is the $n$-th symmetric power of a Hilbert space $\cH$,
see \S 2 of \cite{BC}. The $C^*$-algebra $C^*(\cK)$ is generated by
isometries $\mu_K$, for $K\in \cP_\cK$, with $\mu_K^* \mu_K=1$,
and such that, for $K=K_1\#\cdots \# K_n$, $\mu_K=\mu_{K_1}\cdots \mu_{K_n}$.
The $C^*$-algebra $C^*(\cK)$ is an infinite tensor product of Toeplitz algebras
$C^*(\cK)=\otimes_{K\in \cP} \tau_K$.

\smallskip

Let $\lambda: \cK \to \N$ be a knot invariant that behaves multiplicatively
under connected sums, $\lambda(K_1\# K_2)=\lambda(K_1) \lambda(K_2)$.
Any such invariant determines a semigroup homomorphism $\lambda:(\cK,\#)
\to (\N,\cdot)$.

\smallskip

\begin{ex}\label{Zknotinvprod} {\rm
The Alexander polynomial $\Delta_K(t)\in \Z[t,t^{-1}]$ of a knot $K$ is
multiplicative under connected sums. Thus, for instance, setting $\lambda(K)$ to be
the absolute value of the coefficient of the top degree term of $\Delta_K(t)$ provides 
an example of such a semigroup homomorphism $\lambda: \cK \to \N$.}
\end{ex}

\smallskip

Simpler examples can be obtained by considering additive invariants.
Let $\kappa: \cK \to \Z_+$ be a non-negative integer invariant of knots
satisfying $\kappa(K_1\# K_2)= \kappa(K_1) + \kappa(K_2)$. For a
choice of a positive integer $q\in \N$ (for example, $q=2$), the invariant
$\lambda(K)=q^{\kappa(K)}$ satisfies the multiplicative property as above.

\smallskip

\begin{ex}\label{addinv}{\em 
There are several examples of knot invariants
with values in non-negative integers that behave additively under connected
sums: for example, the knot genus $g(K)$ satisfy additivity
$g(K_1\# K_2)= g(K_1) + g(K_2)$.}
\end{ex}

\smallskip
\subsection{Alternating knots, crossing number, and genus}

A more interesting example is the crossing number $Cr(K)$, the minimum
number of crossings over all planar diagrams $D(K)$. While it is clear that
$Cr(K_1\# K_2)\leq Cr(K_1) + Cr(K_2)$, it is an open conjecture
that the crossing number is in fact additive, 
$Cr(K_1\# K_2)=Cr(K_1) + Cr(K_2)$. It is known that additivity
is satisfied for alternating knots \cite{Schu}, and for certain classes of knots,
like connected sums of torus knots. A larger class of knots on which
additivity is satisfied is identified in \cite{Diao}. 
Thus, we can either use $Cr(K)$
on the entire semigroup $\cK$, conditionally, or restrict to a subsemigroup
$\cK_a$ of alternating knots, or $\cK_t$ generated by those prime knots 
that are torus knots, or one corresponding to the subclass of \cite{Diao}.

\smallskip

\begin{thm}\label{ZetaCrg}
Let $\cP_{\cK,a}\subset \cP_{\cK}$ be the set of prime knots that are
alternating, and consider the bosonic Fock space 
$\ell^2(\cK_a)=\oplus_n S^n \ell^2(\cP_{\cK,a})$.  The $C^*$-algebra
$C^*(\cK_a)=\otimes_{K\in \cP_{\cK,a}} \tau_K$ acts by bounded operators
on the Hilbert space $\ell^2(\cK_a)$, with $\mu_K \epsilon_{K'}=\epsilon_{K\# K'}$.
For a fixed $q\in \N$, with $q\geq 2$, and for all $t\in \R$, setting $\sigma_t(\mu_K)=
q^{it (Cr(K)+g(K))} \mu_K$ defines a time evolution $\sigma: \R \to {\rm Aut}(C^*(\cK_a))$,
with Hamiltonian $H \epsilon_K = (Cr(K)+g(K)) \log(q)\, \epsilon_K$. The partition
function is given by the series
\begin{equation}\label{seriesZ}
 Z_a(\beta)=\Tr(e^{-\beta H}) =\sum_{K\in \cK_a} q^{-\beta (Cr(K)+g(K))} 
\end{equation} 
converges in the range $\beta \geq \beta_+ =\log\frac{2^{20}}{3^6} - 6 \log\log2$ 
and diverges for $\beta < \beta_-$, where $\beta=\beta_-$ is the unique solution of
$$ \beta - 6 \log\left( \frac{q^{-\beta}}{1-q^{-\beta}}\right) = 2 \log(20) - 6 \log \log 2, $$
with $\beta_-=\beta_-(q)\leq 1.9391\cdots$ for all $q\in \N$ with $q\geq 2$.
\end{thm}

\proof The adjoint $\mu_K^*$ acts as $\mu_K^* \epsilon_{K'}=0$ of $K$ does not divide $K'$ 
in the semigroup $(\cK_a,\#)$ and $\mu_K^* \epsilon_{K'}=\epsilon_{K''}$ if $K'=K\# K''$ in $\cK_a$.
These satisfy the relation $\mu_K^*\mu_K=1$, while $\mu_K \mu_K^*$ is the orthogonal
projection on the subspace of $\ell^2(\cK_a)$ generated by all $K'$ that are divisible by 
$K$ in $(\cK_a, \#)$. Thus, setting $\mu_K \epsilon_{K'}=\epsilon_{K\# K'}$ determines a
representation of $C^*(\cK_a)$ on $\ell^2(\cK_a)$. For $K=K_1\# K_2$, we have $\mu_K=\mu_{K_1}\mu_{K_2}$ $C^*(\cK_a)$ and the time evolution satisfies
$\sigma_t(\mu_K)=q^{it (Cr(K)+g(K))} \mu_K=q^{it (Cr(K_1)+g(K_1))} q^{it (Cr(K_2)+g(K_2))} 
\mu_{K_1}  \mu_{K_2} = \sigma_t(\mu_{K_1}) \sigma_t(\mu_{K_2})$, since both $Cr$ and
the genus are additive on connected sums of alternating knots.
It also clearly satisfies $\sigma_{t+s}(X)=\sigma_t (\sigma_s(X))$ for $X\in C^*(\cK_a)$ 
and for all $t,s\in \R$. Thus, the time evolution is indeed a $1$-parameter family of
automorphisms of the algebra, that is, a group homomorphism 
$\sigma: \R \to {\rm Aut}(C^*(\cK_a))$. The Hamiltonian $H$ is determined (up to an arbitrary
additive constant) by the covariance relation $R(\sigma_t(X))=e^{it H} R(X) e^{-itH}$,
for all $X\in C^*(\cK_a)$ and all $t\in \R$, where $R: C^*(\cK_a)\to \cB(\ell^2(\cK_a))$ is 
the representation described above. The densely defined self-adjoint
unbounded operator defined by $H \epsilon_K = (Cr(K)+g(K)) \log(q)\, \epsilon_K$
satisfies
\begin{eqnarray*}
 e^{it H} R(\mu_K) e^{-itH} \epsilon_{K'}& = & q^{-it (Cr(K')+g(K'))} e^{it H} \epsilon_{K\# K'} \\
 & = & q^{-it (Cr(K')+g(K'))} q^{-it (Cr(K\# K')+g(K\# K'))} \epsilon_{K\# K'} \\ & = &
q^{it (Cr(K)+g(K))} R(\mu_K) \epsilon_{K'} \\ &= & R(\sigma_t(\mu_K)) \epsilon_{K'}. 
\end{eqnarray*}
We have 
$$ \Tr(e^{-\beta H}) =\sum_{K\in \cK_a} \langle \epsilon_K, e^{-\beta H} \epsilon_K \rangle
= \sum_{K\in \cK_a} q^{-\beta (Cr(K)+g(K))} $$
$$ = \sum_{n=0}^\infty \sum_{g=0}^\infty N_{n,g} \, q^{-\beta (n+g)}, $$
where $N_{n,g}$ is the number of alternating knots $K$ with $Cr(K)=n$
and $g(K)=g$.  It was shown in Corollary 3.1 of \cite{Stoim} that 
$$ N_{n,g} = O(n^{p_g}), \ \ \ \text{ for } n\to \infty, $$
for some $p_g\in \N$. A more precise estimate is given in Theorem 1.2 of \cite{StoTcherVdo} 
and in Theorem 1.1 of \cite{StoVdo}, which show that
$$ N_{n,g} \sim C_g \, n^{6g-4} \ \ \ \text{ for } n\to \infty, $$
where the $a_n \sim b_n$ means that $a_n/b_n \to 1$ for $n\to\infty$. The
behavior of $C_g$ when $g\to \infty$ can be estimated from below and above
by expressions of the form $\frac{C^g}{(6g)!}$, for constants $C>0$, see Theorem 1.1
of \cite{StoVdo} for a more precise statement. 
We first consider the summation in the crossing number $Cr(K)=n$, for a fixed genus
$g(K)=g$, that is, the series
$$ 1+\sum_{n=1}^\infty N_{n,g} \, q^{-\beta n}. $$
Using the estimate above, the behavior of this series is controlled by that of the polylogarithm
series
$$ {\rm Li}_{4-6g}(q^{-\beta})= \sum_{n=1}^\infty n^{6g-4} q^{-\beta n}, $$
which converges for all $\beta>0$.  We then consider the summation in the genus $g(K)=g$.
The polylogarithm function satisfies
$$ {\rm Li}_{-m}(z)= (z\frac{\partial}{\partial z})^m \frac{z}{1-z}=\sum\limits_{k=0}^m 
k!\,S(m+1,k+1)\,(\frac{z}{1-z})^{k+1} $$
$$ = \frac{1}{(1-z)^{m+1}} \sum_{k=0}^{m-1} \large\langle \begin{array}{c} m \\ k 
\end{array}\large\rangle z^{m-k}, $$
where $S(a,b)$ are the Stirling numbers of the second kind
$$ S(a,b) =\frac{1}{b!} \sum_{j=0}^b (-1)^{b-j} \binom{b}{j} j^a, $$
while
$$ \large\langle \begin{array}{c} m \\ k 
\end{array}\large\rangle = \sum_{j=0}^{k+1} (-1)^j \binom{m+1}{j} (k-j+1)^m $$
are the Eulerian numbers. 
The Stirling numbers of the second kind have upper and lower bounds of the
form \cite{ReDob}
$$ \frac{1}{2} (b^2+b+2) b^{a-b-1} -1 \leq S(a,b) \leq \frac{1}{2} \binom{a}{b} b^{a-b}, $$
and, for fixed $b$, the asymptotic behavior of $S(a,b)$ for $a\to \infty$ is of the 
form $S(a,b)\sim b^a / b!$. Moreover, the ordered Bell numbers 
$\fb_a =\sum_{b=0}^a b! S(a,b)$ behave for $a\to \infty$ like \cite{Spru}
$$ \fb_a \sim \frac{a!}{2 (\log(2))^{a+1}}. $$
When $q^{-\beta}\leq 1/2$, that is, when $\beta> \frac{\log 2}{\log q}$,  
we have $q^{-\beta} \leq (1-q^{-\beta})$, hence
$$ \left(\frac{q^{-\beta}}{1-q^{-\beta}} \right)^{6g-3} \leq \left(\frac{q^{-\beta}}{1-q^{-\beta}} \right)^{k+1}
\leq \frac{q^{-\beta}}{1-q^{-\beta}}, $$
for all $k=0,\ldots,6g-4$.
Thus, the result of the first summation in $n=Cr(K)$ can be approximated, for large $g=g(K)$, by
upper and lower bounds of the form
$$ {\rm Li}_{4-6g}(q^{-\beta}) \leq \fb_{6g-4} \, \frac{q^{-\beta}}{1-q^{-\beta}} \sim \frac{(6g-4)!}{2 (\log 2)^{6g-4}} \, \frac{q^{-\beta}}{1-q^{-\beta}}  $$
$$  \frac{(6g-4)!}{2 (\log 2)^{6g-4}}  \left( \frac{q^{-\beta}}{1-q^{-\beta}} \right)^{6g-3} \sim 
\fb_{6g-4} \, \left( \frac{q^{-\beta}}{1-q^{-\beta}} \right)^{6g-3}
\leq {\rm Li}_{4-6g}(q^{-\beta}). $$
Then, in this range of values of $\beta$, the series defining the partition function 
$Z(\beta) =\sum_K e^{-\beta H_K}$, with $H_K=\langle \epsilon_K, H \epsilon_K \rangle$,
is controlled from above by the behavior of 
$$ \sum_{g=1}^\infty C_g \, \frac{(6g-4)!}{2 (\log 2)^{6g-4}} \, q^{-\beta g} .$$
Using $C_g \sim \frac{C^g}{ (6g)! }$ we obtain 
\begin{equation}\label{estC}
 C_g \, \frac{(6g-4)!}{2 (\log 2)^{6g-4}} \, q^{-\beta g} \sim
\frac{(\log 2)^4}{2} \, \frac{e^{g (\log C -\beta -6 \log\log 2)}}{(6g-3)(6g-2)
(6g-1) 6g}. 
\end{equation}
When $\beta \geq  \log C - 6 \log\log 2$ the above series converges, with
convergence in the case $\beta =\log C-6 \log\log2$ ensured by 
the polynomial in the denominator.
Thus, in the range $\beta\geq \frac{\log 2}{\log q}$, 
the partition function $Z(\beta)=\Tr(e^{-\beta H})$ converges for all
$$ \beta \geq \max\{ \frac{\log 2}{\log q}, \log C-6 \log\log2 \}. $$
 On the other hand, in this same range, the series defining the
partition function is controlled from below by a series of the form
$$ \sum_{g=1}^\infty C_g \, \frac{(6g-4)!}{2 (\log 2)^{6g-4}} \, \lambda_\beta^{6g-3} q^{-\beta g} , $$
where $\lambda_\beta =q^{-\beta}/(1-q^{-\beta})$. In this case we have
\begin{equation}\label{estClambdabeta}
 C_g \, \frac{(6g-4)!}{2 (\log 2)^{6g-4}} \, \lambda_\beta^{6g-3} \, q^{-\beta g} \sim
\frac{(\log 2)^4}{2 \lambda_\beta^3} \, \frac{e^{g(\log C -\beta -6 \log\log 2 +6 \log \lambda_\beta)}}
{(6g-3)(6g-2)(6g-1) 6g}. 
\end{equation}
The corresponding series converges for $\beta - 6 \log\lambda_\beta \geq \log C -6 \log\log 2$
and diverges for $\beta - 6 \log\lambda_\beta < \log C -6 \log\log 2$. Notice that, since in
this range we have $\lambda_\beta \leq 1$, the convergence condition 
$\beta \geq  \log C - 6 \log\log 2$ for the upper bound also implies this convergence,
as it should, while the divergence condition $\beta - 6 \log\lambda_\beta < \log C -6 \log\log 2$
gives a range of divergence for the series defining the partition function $Z(\beta)$: we have
divergence for 
$$ \frac{\log 2}{\log q}\leq \beta < 6 \log \lambda_\beta +\log C -6 \log\log 2. $$
Consider then the case where $\beta < \frac{\log 2}{\log q}$. In this case we have
$q^{-\beta} > (1-q^{-\beta})$, that is, $\lambda_\beta >1$, and, for all $k=0,\ldots,6g-4$,
$$ 
\frac{q^{-\beta}}{1-q^{-\beta}} \leq \left(\frac{q^{-\beta}}{1-q^{-\beta}} \right)^{k+1}
\leq \left(\frac{q^{-\beta}}{1-q^{-\beta}} \right)^{6g-3}.
$$
In this case, the result of the first summation can be approximated from above, for large $g=g(K)$, with
$$ {\rm Li}_{4-6g}(q^{-\beta}) \leq \fb_{6g-4} \, \lambda_\beta^{6g-3} \sim 
\frac{(6g-4)!}{2 (\log 2)^{6g-4}} \, \lambda_\beta^{6g-3} , $$
 and from below with 
$$  \frac{(6g-4)!}{2 (\log 2)^{6g-4}} \, \lambda_\beta \sim \fb_{6g-4} \, \lambda_\beta\leq
{\rm Li}_{4-6g}(q^{-\beta}). $$
Thus, in this case, the series that determines the partition function is controlled from above by the 
 behavior of the series
$$ \sum_{g=1}^\infty C_g \frac{(6g-4)!}{2 (\log 2)^{6g-4}} \, \lambda_\beta^{6g-3}. $$
As above, we can estimate this with \eqref{estClambdabeta}. Again, the resulting
series converges for $\beta - 6 \log \lambda_\beta \geq \log C - 6 \log\log 2$. Here
$\lambda_\beta >1$, so this inequality also implies the inequality $\beta \geq 
\log C - 6 \log\log 2$, which in this case gives the convergence of the lower bound,
here of the form \eqref{estC}. The divergence of the lower bound happens for
$\beta < \log C - 6 \log\log 2$. Thus, in the range $\beta < \frac{\log 2}{\log q}$ we
have convergence when 
$$  6 \log \lambda_\beta + \log C - 6 \log\log 2 \leq  \beta < \frac{\log 2}{\log q}, $$
and divergence for
$$ \beta < \min\{ \frac{\log 2}{\log q}, \log C - 6 \log\log 2 \} . $$
As in Theorem 1.1 of \cite{StoVdo}, we can take the constant $C$
to be $C=400$ for the lower bound on $C_g$ and $C=2^{20}/3^6 \sim 1438.38$
for the upper bound on $C_g$. Using these values we can estimate that the
series $Z(\beta)=\sum_K \langle \epsilon_K, e^{-\beta H} \epsilon_K \rangle$
defining the partition function converges for 
$$ \beta \geq \max\{ \frac{\log 2}{\log q}, \log\frac{2^{20}}{3^6} - 6 \log\log2 \} $$
and for
$$ \beta - 6 \log\lambda_\beta \geq \log\frac{2^{20}}{3^6} - 6 \log\log 2 \ \ \ \text{ and }
\ \ \ \beta < \frac{\log 2}{\log q} $$
while it diverges for
$$ \beta < \min\{ \frac{\log 2}{\log q}, 2 \log(20) - 6 \log\log 2 \} $$
and for
$$ \beta - 6 \log\lambda_\beta < 2\log(20) - 6 \log\log 2 \ \ \ \text{ and }
\ \ \  \beta\geq \frac{\log 2}{\log q}. $$
Consider the condition that the integer $q\in \N$ satisfies
$$ \frac{\log 2}{\log q} < 2 \log 20 -6 \log\log 2. $$
We have $\log 2 =( 2 \log 20 -6 \log\log 2)\log(x)$ for $x\sim 1.0883$, hence
for all $q\in \N$ with $q\geq 2$ the condition above is satisfied.
Then the convergence range above reduces to just the first condition
$\beta \geq \log\frac{2^{20}}{3^6} - 6 \log\log2$, since in the second case the
conditions $\beta< \frac{\log 2}{\log q}$ and $\beta \geq 
\beta - 6 \log\lambda_\beta \geq \log\frac{2^{20}}{3^6} - 6 \log\log2$ cannot
be simultaneously realized since $\frac{\log 2}{\log q} < \log\frac{2^{20}}{3^6} - 6 \log\log2$.
Let $\beta_+:=\log\frac{2^{20}}{3^6} - 6 \log\log2$. 
Similarly, the estimate of the range of divergence gives $\beta<\frac{\log 2}{\log q}$
or $\frac{\log 2}{\log q} \leq \beta <  6 \log\lambda_\beta + 2 \log 20 -6 \log\log 2$.
Note that, in the range $\beta \geq \frac{\log 2}{\log q}$, the function
$\beta - 6 \log\lambda_\beta$ is non-negative and monotonically increasing, with a zero
at $\beta = \frac{\log 2}{\log q}$. Let $\beta_-$ be the unique value of $\beta$ where 
$\beta - 6 \log\lambda_\beta=2 \log 20 -6 \log\log 2\sim 8.1905$. The dependence on $q$ 
of $\beta_-=\beta_-(q)$ is monotonically decreasing, with $\beta_-(q=2)\sim 1.9391$, and with
for example $\lambda_-(q=10^2)\sim 0.3362$ and $\lambda_-(q=10^3)\sim 0.2262$.
Then we obtain that the series defining the partition function is divergent in the range
$\beta < \beta_-$. Note that the function $F(q)=\beta_+ -6 \log\lambda_{\beta_+}(q) - (2 \log 20 -6 \log\log 2)$
is monotonically increasing in the variable $q$, with $F(2)\sim 40.6574$, hence $\beta_- < \beta_+$.
Summarizing, we conclude that, for any choice of $q\in \N$ with $q\geq 2$, the
series defining the partition function $Z(\beta)$ is convergent for $\beta\geq \beta_+$
and divergent for $\beta < \beta_-$.
\endproof

\smallskip

\begin{rem}\label{between}{\rm
The approximation method we used here, based on the estimates of \cite{StoVdo}, 
does not give information on the
behavior of the series defining the partition function in the range
$\beta_-  \leq \beta  < \beta_+ $, but it is reasonable to expect that there will be a point 
$\beta_c\in [\beta_-,\beta_+]$ where a phase transition occurs, so that the series defining
the partition function converges for all $\beta > \beta_c$ and
diverges for all $\beta< \beta_c$. }
\end{rem}

\smallskip

\begin{lem}\label{EulProd}
In the range $\beta\geq \beta_+$, where the series \eqref{seriesZ} 
is convergent, the partition function $Z(\beta)$ has an Euler product
expansion
\begin{equation}\label{EulProdZ}
Z_a(\beta)= \prod_{K\in \cP_{\cK,a}} (1-q^{-\beta (Cr(K)+g(K))})^{-1} .
\end{equation}
\end{lem}

\proof This is a general fact about bosonic Fock spaces and the
trace and determinant of operators. As in \S 2 of \cite{BC}, we
identify $\ell^2(\cK_a)=S \ell^2(\cP_{\cK,a}):=\oplus_{n=0}^\infty S^n \ell^2(\cP_{\cK,a})$,
the bosonic Fock space given by the sum of the symmetric powers of $\ell^2(\cP_{\cK,a})$.
Let $T$ be the densely defined operator on $\ell^2(\cP_{\cK,a})$ with
$T \epsilon_K = q^{-\beta (Cr(K)+g(K))} \epsilon_K$, and let $ST$ be the induced
densely defined operator on the Fock space $\ell^2(\cK_a)$. On a basis
element $\epsilon_{K_1\#\cdots\# K_m}=\epsilon_{K_1}\cdots \epsilon_{K_m}$,
this satisfies $ST \epsilon_{K_1\#\cdots\# K_m}=q^{-\beta (Cr(K_1)+g(K_1))} \cdots
q^{-\beta (Cr(K_m)+g(K_m))} \epsilon_{K_1\#\cdots\# K_m}$. Thus, when the trace
of $ST$ is finite it satisfies
$$ \Tr(ST)=\frac{1}{\det(1-T)}. $$
By direct inspection, we see that $ST=e^{-\beta H}$ and that
$1/\det(1-T)$ is the Euler product of \eqref{EulProdZ}. 
\endproof

\smallskip
\subsection{Statistical mechanics of knots without interaction}

We then have, for the $C^*$-dynamical system $(C^*(\cK_a),\sigma_t)$ described above, the
analog of Proposition 8 of \cite{BC}. 

\smallskip

\begin{prop}\label{KMSKa}
For every $\beta>0$ there is a unique KMS$_\beta$ state for $(C^*(\cK_a),\sigma_t)$,
which is the infinite tensor product of unique KMS$_\beta$ states $\phi_{\beta, K}$
for $K\in \cP_{\cK,a}$, on the Toeplitz algebra $\tau_K$ with the induced time evolution,
with eigenvalue list 
\begin{equation}\label{eigenlist}
 \Sigma(\phi_{\beta,K})=\{ (1-q^{-\beta (Cr(K)+g(K))}) q^{-\beta n (Cr(K)+g(K))} \}_{n\in \N}. 
\end{equation} 
For $\beta\geq \beta_+$ the KMS state is a Gibbs state of the form
$$ \phi_\beta (X)=\frac{1}{Z(\beta)} \Tr( X e^{-\beta H} ), \ \ \  \forall X \in C^*(\cK_a), $$
while for $\beta < \beta_-$ the KMS state is of type III.
\end{prop}

\proof On the Toeplitz algebra $\tau_K$, for some $K\in \cP_{\cK,a}$, the induced
time evolution is determined by $\sigma_t(\mu_K) = q^{it (Cr(K)+g(K))} \mu_K$.
A KMS$_\beta$ state on $(\tau_K, \sigma_t)$ will necessarily vanish on all eigenvectors
of the time evolution with $\sigma_t(X)=\lambda^{it} X$ where $\lambda\neq 1$, while by
the KMS condition it will satisfy $$\varphi_{\beta,K}(\mu_K \mu_K^*)=
\varphi_{\beta,K}(\mu_K^* \sigma_{i\beta}(\mu_K))=q^{-\beta (Cr(K)+g(K))} 
\varphi_{\beta,K}(\mu_K^* \mu_K) $$ $$ =q^{-\beta (Cr(K)+g(K))} \varphi_{\beta,K}(1)=
q^{-\beta (Cr(K)+g(K))}. $$
The complementary projection $1-\mu_K \mu_K^*$ then has $\varphi_{\beta,K}(1-\mu_K \mu_K^*)=
1-q^{-\beta (Cr(K)+g(K))}$. On powers $\mu_K^n (\mu_K^*)^m$ the KMS state vanishes unless $n=m$,
in which case $\varphi_{\beta,K}(\mu_K^n (\mu_K^*)^n)=q^{-\beta n(Cr(K)+g(K))}$, by the same
argument. Note that, since we are working with alternating knots $n (Cr(K)+g(K))=
Cr(K\#\cdots \# K)+ g(K\# \cdots \# K)$, with the connected sum taken $n$ times. 
The same argument used in Proposition 8 of \cite{BC} then shows that this determines
uniquely the KMS$_\beta$ state $\phi_{\beta,K}$, and the fact that this implies the
uniqueness of the KMS$_\beta$ state on the tensor product $C^*$-algebra $C^*(\cK_a)=
\otimes_{K\in \cP_{\cK,a}} \tau_K$. As in case (b) of Proposition 8 of \cite{BC} the finiteness
of $Z(\beta)=\Tr(e^{-\beta H})$ for $\beta\geq \beta_+$ shows that the KMS$_\beta$ state
is of the Gibbs form (by uniqueness, since the Gibbs state is clearly a KMS$_\beta$ state).
In the range $\beta< \beta_-$ where the series defining the partition function is
divergent, one uses the same argument used in \cite{BC}, based on the result of \cite{AraWoods}.
Namely, as in Lemma 2.14 of \cite{AraWoods}, if $\{ \lambda_{\nu,i} \}$ is the
eigenvalue list of an infinite tensor product $M=\otimes_\nu M_\nu$
of type I factors, then $M$ is of type I if and only if $\sum_\nu |1-\lambda_{\nu 1}|<\infty$;
of type II if and only if $n_\nu<\infty$ for all $\nu$ and 
$\sum_{\nu, i} |n_\nu^{-1/2}-\lambda_{\nu i}^{1/2}|^2 <\infty$; and, when $\lambda_{\nu, 1}\geq \delta$
for some $\delta$ for all $\nu$, $M$ is of type III if and only if $$\sum_{\nu, i} \lambda_{\nu, i} \inf\{ | \frac{\lambda_{\nu 1}}{\lambda_{\nu i}} -1 |^2, C\} =\infty$$ for some (hence all) $C>0$. In our
case, with the eigenvalue list \eqref{eigenlist}, we have
$\lambda_{\nu,1}=1-q^{-\beta(Cr(K)+g(K))}$ hence $|1-\lambda_{\nu,1}|=
q^{-\beta(Cr(K)+g(K))}$. In the range $\beta<\beta_-$ the series $\sum_K q^{-\beta(Cr(K)+g(K))}$
is divergent, hence type I is excluded. Similarly, type II is excluded because $n_\nu=\infty$.
For a fixed $\beta$, the condition $\lambda_{\nu, 1}\geq \delta$ is satisfied with
$\delta = 1-q^{-\beta}$, and we have
$$ \sum_{\nu, j} \lambda_{\nu, j} \inf\{ | \frac{\lambda_{\nu 1}}{\lambda_{\nu j}} -1 |^2, C\} 
\sim  \sum_{K,j} (1-q^{-\beta (Cr(K)+g(K))}) q^{-\beta j (Cr(K)+g(K))} =\infty $$
hence the factor is type III. 
\endproof

\begin{rem}\label{noDix}{\rm Notice that, since we do not have in this
case a complete analysis of the behavior of the partition function in
the intermediate range $\beta_-\leq \beta < \beta_+$, we do not have
in this case the direct analog of case (c) of Proposition 8 of \cite{BC}.}
\end{rem}

\smallskip

\begin{lem}\label{tildebetamin}
For a fixed $q\in \N$, $q\geq 2$, 
there is a unique solution $\tilde\beta_-=\tilde\beta_-(q)$, with
$\tilde \beta_- > \frac{\log 2}{\log q}$, to the equation
\begin{equation}\label{eqtildebeta}
\beta - 6 \log \lambda_\beta + 6 \log \beta = \log C - 6 \log \log q,
\end{equation}
where $C=400$ and
$$ \lambda_{\beta} = \frac{q^{-\beta}}{1-q^{-\beta}}. $$
The value $\tilde\beta_-(q)$ satisfies $\tilde\beta_-(q) < \beta_-(q)$,
where $\beta_-(q)$ is, as in Theorem \ref{ZetaCrg}, the
unique solution of $\beta -6 \log \lambda_\beta =
  \log C - 6 \log\log 2$. 
\end{lem}

\begin{figure}
\begin{center}
\includegraphics[scale=0.5]{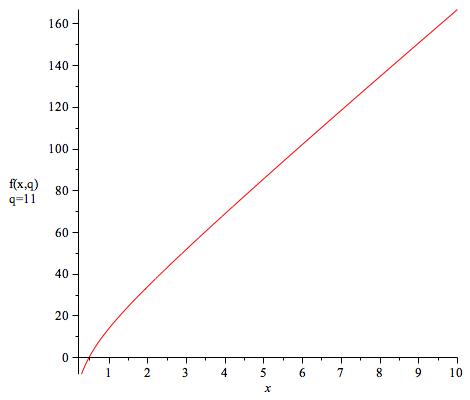}
\end{center}
\caption{The function $f(\beta,q)$ for $\beta > \frac{\log2}{\log q}$ and $q=11$.\label{Figfbetaq}}
\end{figure}

\proof For $\beta= \log 2 /\log q$ we have
$(\beta -6 \log \lambda_\beta)|_{\beta =\frac{\log 2}{\log q}}=\frac{\log 2}{\log q}$
hence $$(\beta - 6 \log \lambda_\beta + 6 \log \beta)|_{\beta =\frac{\log 2}{\log q}} =
\frac{\log 2}{\log q} + 6 \log\log 2 - 6 \log \log q < \log C - 6 \log \log q, $$
since we have seen in Theorem \ref{ZetaCrg} that, for all $q\in \N$ with $q\geq 2$,
\begin{equation}\label{log2qC}
 \frac{\log 2}{\log q} < \log C - 6 \log\log 2. 
\end{equation} 
For $\beta\geq \log 2/\log q$ the function $f(\beta,q):=\beta - 6 \log \lambda_\beta + 6 \log \beta$
is monotonically increasing and unbounded for $\beta \to \infty$ (see the plot in Figure \ref{Figfbetaq}, 
hence there will be a unique $\tilde\beta_-=\tilde\beta_-(q)$ where \eqref{eqtildebeta} holds. Finally,
we see that at $\beta =\beta_-(q)$ we have
$$ (\beta -6 \log \lambda_\beta+ 6 \log \beta)|_{\beta =\beta_-(q)} =
 \log C - 6 \log\log 2 + 6 \log \beta_-(q). $$
Notice that we have $\beta_-(q) > \log 2/ \log q$ because of \eqref{log2qC},
hence we find
$\beta_- -6 \log \lambda_{\beta_-}+ 6 \log \beta_- > \log C - 6 \log \log q$,
hence $\beta_-(q)> \tilde\beta_-(q)$.
\endproof

\smallskip

Thus, the range $\beta< \tilde\beta_-(q)$ is
contained in the range of divergence of the series defining the partition function,
as we have seen in Theorem \ref{ZetaCrg}.

\smallskip

\begin{thm}\label{IIIq}
Let $\tilde\beta_-=\tilde\beta_-(q)$ be as in Lemma \ref{tildebetamin}.
For $\beta < \tilde\beta_-(q)$, the
unique KMS$_\beta$ state is of type III$_{q^{-\beta}}$.
\end{thm}

\proof 
The argument is similar to Lemma 4.5.1 of \cite{Jacob} and 
Lemma 2.4 of \cite{NeRu}. We need to show that $q^{-\beta}$ belongs
to the asymptotic ratio set, see Definition 3.2 and Lemma 3.6 of \cite{AraWoods}. 
As in Lemma 4.5.1 of \cite{Jacob}, for given $\beta$, let $N\in \N$ be chosen 
so that $\beta N > \beta_+$ and, for a chosen $K\in \cP_{\cK,a}$, consider
the projector $e_K=1-\mu_K^N (\mu_K^*)^N$ in the Toeplitz algebra $\tau_K$,
and the projection $e=\prod_{K\in \cP_{\cK,a}} e_K$, as weak limit of projections
in the tensor product von Neumann algebra. Since $\beta N > \beta_+$ we have,
using the Euler product of Lemma \ref{EulProd}, 
$$ \phi_\beta(e)=\prod_{K\in \cP_{\cK,a}} (1-q^{-\beta N (Cr(K)+g(K))}) =Z(\beta N)^{-1}\neq 0, $$
hence $e\neq 0$. Setting $\tilde\phi_{\beta,e}(X)=\phi_\beta(X)/\phi_\beta(e)$ determines
a KMS state on the compression of the algebra with the projection $e$. For each prime knot
$K\in \cP_{\cK,a}$ we similarly have $\tilde\phi_{\beta,e,K}(X) = \phi_{\beta, K}(X) (1-q^{-\beta N (Cr(K)+g(K))})^{-1}$. The eigenvalue list of $\tilde\phi_{\beta,e,K}$ is then
$$ \Sigma(\tilde\phi_{\beta,e,K})=\{ \frac{(1-q^{-\beta (Cr(K)+g(K))}) q^{-\beta n (Cr(K)+g(K))}}{(1-q^{-\beta N (Cr(K)+g(K))})} \}_{n\in \N}. $$
By the results of \cite{StoVdo}, the number of knots $K$ in $\cK_a$ with a given value
$Cr(K)+g(K)=n$ is given by 
\begin{equation}\label{Nnest}
 N(n)=\#\{ K\in \cK_a\,|\, Cr(K)+g(K)=n\} \sim \sum_{g=1}^n \frac{C^g}{(6g)!} (n-g+1)^{6g-4}. 
\end{equation} 
Thus, we have $N(n)$ knots $K_1,\ldots, K_{N(n)}$ for which 
$q^{\beta(Cr(K_i)+g(K_i))}=q^{-\beta n}$. Consider two disjoint sets $\cX_1(n)=\{ K_1,\cdots, K_{N(2n)} \}$,
the set of knots with $q^{\beta(Cr(K_i)+g(K_i))}=q^{-\beta 2n}$ and a subset $\cX_2(n)=\{ K'_1,\ldots, 
K'_{N(2n)} \}$ of the same cardinality of the set of knots with $q^{\beta(Cr(K_i)+g(K_i))}=q^{-\beta (2n+1)}$.
Consider the set  $\cF_n$ of functions from the set $\cX(n)=\cX_1(n)\cup \cX_2(n)$ to the
set $\underline{N}=\{ 0, \ldots, , N-1\}$, namely $\cF_n=\cF(\cX(n), \underline{N})$. In 
this set, consider the delta functions $\delta_{K_i}$ and $\delta_{K'_i}$ for $i=1,\ldots, N(2n)$.
Setting
$$ \mu(f)=\prod_{i=1}^{N(2n)} \lambda_{K_i,f(K_i)} \lambda_{K'_i, f(K'_i)}, $$
where 
$$ \lambda_{K_i,j}=\frac{(1-q^{2n \beta})}{(1-q^{-2nN\beta})} q^{-\beta 2n j}, $$
defines a measure on the set $\cF_n$. This satisfies
$$ \mu(\delta_{K_i}) = \mu(\delta_{K_1}) = 
\left( \frac{(1-q^{2n \beta})}{(1-q^{-2nN\beta})} \right)^{N(2n)} \cdot
\left( \frac{(1-q^{(2n+1) \beta})}{(1-q^{-(2n+1) N\beta})} \right)^{N(2n)}\cdot q^{-\beta 2n} =:\mu(n). $$
The measure of the set $\{ \delta_{K_i} \}$ is equal to $\mu(\{ \delta_{K_i} \})=N(2n) \mu(n)$.

\smallskip

By \eqref{Nnest}, the behavior of the series $\sum_n N(2n) \mu(n)$ can be estimated in
terms of the behavior of 
$$ \sum_n N(2n) q^{-\beta 2n} = \sum_n \sum_{k+g=2n} N_{k,g} q^{-\beta (k+g)}, $$
where $N_{k,g}$ is the number of alternating knots with $Cr(K)=k$ and $g(K)=g$.
Note that this is a subseries of the series $\sum_g \sum_k N_{k,g} q^{-\beta (k+g)}$, 
whose behavior we analyzed in Theorem \ref{ZetaCrg}.
In particular, we know that for $\beta < \beta_-$ the series $\sum_g \sum_k N_{k,g} 
q^{-\beta (k+g)}$ diverges. We now need to check whether the subseries corresponding
to the terms with $k+g$ even also diverges. We first show that we can express
this series in terms of the Lerch transcendents, replacing the polylogarithms
used in the case of the full series in Theorem \ref{ZetaCrg}.

\smallskip

 Let $\Phi(z,s,\alpha)$ be the Lerch transcendent
 \begin{equation}\label{Lerch}
 \Phi(z,s,\alpha) = \sum_{\ell\geq 0} \frac{z^\ell}{(\alpha +\ell)^s}.
 \end{equation}
 We can then write the series above as 
 $$ \sum_{k,g\geq 0\,:\, k+g\, {\rm even}} N_{k,g} q^{-\beta (k+g)} \sim
 \sum_{g\geq 0} q^{-2\beta g} \frac{C^g}{(6g)!} \sum_{\ell\geq 0} q^{-2\beta \ell} (g+2\ell)^{6g-4} $$
 \begin{equation}\label{seriesLerch}
  = \sum_{g\geq 0} \frac{C^g 2^{6g-4} q^{-2\beta g}}{(6g)!} \,\, 
 \Phi(q^{-2\beta}, 4-6g, \frac{g}{2}). 
 \end{equation}
 
 \smallskip
 
\begin{figure}
\begin{center}
\includegraphics[scale=0.6]{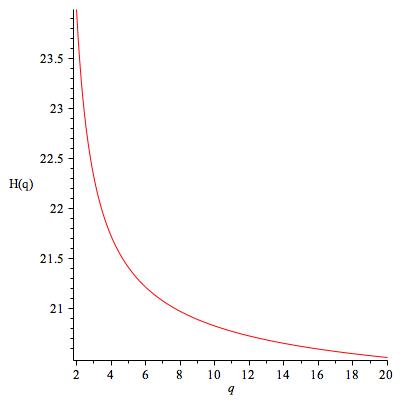}
\end{center}
\caption{The function $H(q)$ for $q\geq 2$.\label{FigHq}}
\end{figure}

  The Lerch transcendent $\Phi(z,s,\alpha)$ has a Taylor expansion
  \begin{equation}\label{LerchTaylor}
  \Phi(z,s,\alpha)= z^{-\alpha} \left( \Gamma(1-s) (-\log(z))^{s-1} + \sum_{j\geq 0} \zeta(s-j,\alpha)
  \frac{\log^j(z)}{j!} \right),
  \end{equation}
  which is valid for $|\log(z)|< 2\pi$, $s\notin \N$ and $\alpha \notin \Z_{\leq 0}$. In our setting
  we have $z=q^{-2\beta}$, hence $|\log(z)|=2\beta \log(q)$. 
  One can check that the function 
  $$ H(q): = ( \beta - 6 \log \lambda_\beta + 6 \log \beta)|_{\beta = \frac{\pi}{\log q}} - (\log C - 6 \log \log q) $$
  is positive for $q\geq 2$ (see the plot in Figure \ref{FigHq}), hence $\frac{\pi}{\log q} > \tilde\beta_-(q)$.
  Thus, in the range $\beta < \tilde\beta_-(q)$ the Taylor expansion above applies. 
  
  \smallskip
  
  Then we have
  $$ \Phi(q^{-2\beta}, 4-6g, \frac{g}{2}) = q^{\beta g} \left( \frac{(6g-4)!}{(2\beta \log(q))^{6g-3}} +
  \sum_{j\geq 0} \zeta(4-6g-j,\frac{g}{2}) \frac{(-2\beta \log(q))^j}{j!} \right). $$
  Thus, the general term of series above has a leading contribution of the form
  \begin{equation}\label{Lerchleading}
    \frac{C^g q^{-\beta g}}{(6g)(6g-1)(6g-2)(6g-3)} \, \cdot \, \frac{2^{6g-4}}{(2\beta \log(q))^{6g-3}} . 
  \end{equation}  
  Notice that this is the analog of the leading term of the form
 $$ \frac{C^g q^{-\beta g}}{(6g)(6g-1)(6g-2)(6g-3)} \, \cdot \,  \frac{1}{2 (\log 2)^{6g-4}}  $$
 for the full series, as in \eqref{estC} of Theorem \ref{ZetaCrg}.
  Arguing in a similar way, we then see that the series 
  $$ \sum_g \frac{C^g q^{-\beta g}}{(6g)(6g-1)(6g-2)(6g-3)2(\beta \log(q))^{6g-3}} $$
  is divergent in the range $\beta < \tilde\beta_-(q)$, hence so is 
  the series \eqref{seriesLerch}.  Thus, we obtain the divergence result
  $$ \sum_{n\geq n_0} N(2n) \mu(n) =\infty. $$

\smallskip 

We then proceed in the same way as in Lemma 4.5.1 of \cite{Jacob}.
The bijection $\Psi_n: \cX_1(n) \to \cX_2(n)$ determines a bijection of the set
of delta functions, and we have
$$ \frac{\mu(\Psi_n(\delta_{K_i}))}{\mu(\delta_{K_i})} = \frac{\lambda_{K_i,0}\lambda_{K'_i,1}}
{\lambda_{K_i,1}\lambda_{K'_i,0}} = \frac{q^{-\beta (2n+1)}}{q^{-\beta 2n}} =q^{-\beta}, $$
which shows that $q^{-\beta}$ is in the asymptotic ratio set. 
\endproof

\smallskip
\subsection{Intermezzo: statistical physics of $\Q^*_+$}\label{interSec}

In preparation for the construction we will illustrate in the following section,
we discuss here a toy model, based on the multiplicative group $\Q^*_+$
and its reduced group algebra $C^*_r(\Q^*_+)$ acting on the Hilbert space
$\ell^2(\Q^*_+)$. Here we regard $\Q^*_+$ as the discrete infinitely generated
abelian group, generated by the primes, $\Q^*_+=\prod_{p\in \cP} p^\Z$.
We want to construct a quantum statistical mechanical system whose
algebra of observables contains $C^*_r(\Q^*_+)$, with Hilbert space of states
$\ell^2(\Q^*_+)$ and 
with a Hamiltonian generator densely defined on $\ell^2(\Q^*_+)$, so that the partition
function $Z(\beta)=\Tr(e^{-\beta H})$ is finite for sufficiently large $\beta>0$. 

\smallskip

It is easy to construct such a system for the multiplicative semigroup $\N$
and the $C^*$-algebra $C^*_r(\N)$ acting on $\ell^2(\N)$. For instance
by considering the ``system without interaction" of \cite{BC} with time
evolution $\sigma_t(\mu_n)=n^{it}\mu_n$ and with $H \epsilon_n =\log(n) \epsilon_n$.
However, the natural extension of this system from the semigroup $\N$ to the group 
$\Q^*_+$ by $\sigma_t(\mu_{a/b}) =(a/b)^{it} \mu_{a/b}$
will no longer satisfy the condition $\Tr(e^{-\beta H})<\infty$ for large $\beta$.

\smallskip

We proceed in a slightly different way, motivated by the analogies between
quantum statistical mechanical systems and spectral triples discussed in
\cite{GMT}. We consider first the case of a single prime $p$ and the group 
$p^\Z\simeq \Z$, and then the case of the group $\Q^*_+$.

\smallskip

Recall that a spectral triple $(\cA,\cH,D)$ is the datum of an
involutive algebra $\cA$, a representation of $\cA$ by bounded operators
on $\cH$ and a self-adjoint operator $D$, densely defined on $\cH$, with
compact resolvent $(D^2+1)^{-1/2}\in \cK$ and such that the commutators
$[D,a]$ with all $a\in \cA$ are bounded operators on $\cH$, see \cite{CoS3}.

\smallskip

\begin{lem}\label{S3p}
Consider the algebra $\C[p^\Z]\subset C^*(p^\Z)\simeq C(S^1)$, acting on $\ell^2(p^\Z)$,
and the operator defined by $D_p \epsilon_{p^n} = n \log(p) \epsilon_{p^n}$.
The datum $(\C[p^\Z],\ell^2(p^\Z),D_p)$ is a spectral triple.
\end{lem}

\proof It is easy to check that all properties are satisfied. We check explicitly
the bounded commutator condition. Let $\delta_{p^m}$ be the operator on $\ell^2(p^\Z)$ 
corresponding to the delta function $\delta_{p^m}\in \C[p^\Z]$. It acts by
$\delta_{p^m} \epsilon_{p^n}=\epsilon_{p^{n+m}}$. Thus, we have
$$ (D_p \delta_{p^m} - \delta_{p^m} D_p) \epsilon_{p^n} =
((m+n) \log(p) -n \log(p)) \epsilon_{p^{m+n}} = m\log(p) \epsilon_{p^{m+n}}, $$
hence the bounded commutator condition is satisfied for all element of the
dense subalgebra $\C[p^\Z]$ of $C^*_r(p^\Z)$. 
\endproof

\smallskip

\begin{rem}\label{noS3Q}{\rm
Notice that, in the case of the group $\Q^*_+$, the operator 
$D$ acting on a basis element $\epsilon_r$ of $\ell^2(\Q^*_+)$ as
$D\,\,\epsilon_r = (n_1 \log(p_1)+\cdots + n_k\log(p_k)) \epsilon_r$,
for $r=p_1^{n_1}\cdots p_k^{n_k} \in \Q^*_+$, with $n_i\in \Z$
has bounded commutators with elements of the dense subalgebra
$\C[\Q^*_+]$ of $C^*_r(\Q^*_+)$, by the same argument of Lemma \ref{S3p}. 
However, $D$ does not have compact
resolvent, since the set $\{ \sum_{i=1}^k n_i \log(p_i)\,|\, n_i\in \Z,\, p_i\in \cP,\, k\in \N\}$
is dense in $\R$, hence $D$ does not determine a spectral triple for $C^*_r(\Q^*_+)$.
}\end{rem}

\smallskip

We now modify the operator above, using the polar
decomposition of the Dirac operator $D_p=|D_p|\, F$, 
with $F$ the sign operator.

\begin{lem}\label{HpTr}
Consider the operator $H_p=|D_p|$ acting on a basis $\epsilon_{p^n}$ of $\ell^2(p^\Z)$ as
\begin{equation}\label{Hp}
H_p\, \epsilon_{p^n} = |n| \log(p) \epsilon_{p^n}.
\end{equation}
The operator $e^{-\beta H_p}$ is trace class for all $\beta >0$ with
\begin{equation}\label{TrHpEuler}
\Tr(e^{-\beta H_p}) = 1 + 2 \sum_{n\in \N} p^{-\beta n} =\frac{(1-p^{-2\beta})}{(1-p^{-\beta})^2} .
\end{equation}
\end{lem}

\proof For all $\beta >0$, we have
$$ \Tr(e^{-\beta H_p}) =\sum_{n\in \Z} \langle \epsilon_{p^n}, e^{-\beta H_p}\epsilon_{p^n}\rangle
= 1 + 2 \sum_{n\in \N} p^{-\beta n} = 1 + \frac{2p^{-\beta}}{(1-p^{-\beta})}  $$
$$ = \frac{1}{1-p^{-\beta}} + \frac{p^{-\beta}}{1-p^{-\beta}} = \frac{1+p^{-\beta}}{1-p^{-\beta}}=
\frac{(1-p^{-2\beta})}{(1-p^{-\beta})^2} $$
where the first term corresponds to $\Ker(H_p)=\C \epsilon_1$.
\endproof

This in turn determines an operator $H$ on $\ell^2(\Q^*_+)$ with the
following properties.

\begin{lem}\label{HQTr}
Consider the operator $H$ acting on the basis elements $\epsilon_r$ of $\ell^2(\Q^*_+)$, 
for $r=p_1^{n_1}\cdots p_k^{n_k} \in \Q^*_+$, with $n_i\in \Z$, as
\begin{equation}\label{HopQ}
H \,\,\epsilon_r = (|n_1| \log(p_1)+\cdots + |n_k| \log(p_k)) \epsilon_r.
\end{equation}
Then for $\beta > 1$ the operator $e^{-\beta H}$ is trace class with  
\begin{equation}\label{TrHQzeta}
 \Tr(e^{-\beta H}) = \frac{\zeta^2(\beta)}{\zeta(2\beta)}, 
\end{equation} 
where $\zeta(\beta)$ is the Riemann zeta function.
\end{lem}

\proof We have $\Ker(H) =\C \epsilon_1$, with $\epsilon_1$ the basis vector of $\ell^2(\Q^*_+)$ 
corresponding to the unit in $\Q^*_+$. The spectrum of $H$ is given by
$\Spec(H)=\{ \log(n)\,|\, n\in \N\}$.  The trace is then computed by
$$ \Tr(e^{-\beta H})= \sum_{r\in \Q^*_+} \langle \epsilon_r , e^{-\beta H} \epsilon_r \rangle =
 \sum_{\lambda \in \Spec(H)} m_\lambda e^{-\beta \lambda}, $$
where $m_\lambda$ is the multiplicity. For $\lambda = n_1 \log(p_1) + \cdots + n_k \log(p_k)=\log(n)$
with $n_i\in \N$ and $p_i\in \cP$, and $n=p_1^{n_1}\cdots p_k^{n_k}$,
the multiplicity is $m_\lambda =2^k$, with $k$ the number
of distinct prime factors in $r=p_1^{\pm n_1}\cdots p_k^{\pm n_k}$, 
since for each $n_i$ we have two choices of $\pm n_i \in \Z$.
Thus, we can rewrite the series computing the partition function as
$$ \Tr(e^{-\beta H})= \sum_{n\in \N} \frac{2^{\omega(n)}}{n^{-\beta}}, $$
where $\omega(n)$ is the number of distinct prime factors of $n\in \N$. It is known by
Theorem 301, p.335 of \cite{HarWri} that this series converges for $\beta >1$ with sum
$$ \sum_{n\in \N} \frac{2^{\omega(n)}}{n^{-\beta}} =\frac{\zeta^2(\beta)}{\zeta(2\beta)}, $$
with $\zeta(\beta)=\sum_{n\in \N} n^{-\beta}$ the Riemann zeta function. This can be
seen easily by the form \eqref{TrHpEuler} of the Euler factors, since we have
$$ \frac{\zeta^2(\beta)}{\zeta(2\beta)} =\prod_p \frac{(1-p^{-2\beta})}{(1-p^{-\beta})^2} =
\prod_p (1 + 2 \sum_{k\in \N} p^{-\beta k}) =\sum_{n\geq 1} 2^{\omega(n)}\,\, n^{-\beta}. $$
\endproof

Consider the algebra of bounded operators $\cB(\ell^2(\Q^*_+))$. The operator $H$
described above determines a time evolution of the form
\begin{equation}\label{sigmaBl2Q}
\sigma_t(T) = e^{it H} T e^{-it H}, \ \ \ \forall t\in \R, \ \ \  \forall T\in \cB(\ell^2(\Q^*_+)),
\end{equation}
with partition function as in \eqref{TrHQzeta}
$$ Z(\beta) =\Tr(e^{-\beta H}) = \frac{\zeta^2(\beta)}{\zeta(2\beta)}. $$

\smallskip

The subalgebra $C^*_r(\Q^*_+)\subset \cB(\ell^2(\Q^*_+))$ is not preserved 
by the time evolution  \eqref{sigmaBl2Q}.
We have the following result, which is analogous to Lemma 5.7 of \cite{GMT}.

\begin{prop}\label{inneractQ}
The smallest $C^*$-subalgebra $\cA\subset \cB(\ell^2(\Q^*_+))$ 
that contains $C^*_r(\Q^*_+)$ is invariant under the time evolution \eqref{sigmaBl2Q}
is generated by $C^*_r(\Q^*_+)$ and by projections 
$$ \Pi_{(k,\ell)} \epsilon_{a/b}=\left\{ \begin{array}{ll} \epsilon_{a/b} & k|a \text{ and } \ell |b\\
0 & \text{otherwise} \end{array}\right. $$
The time evolution \eqref{sigmaBl2Q} acts on the algebra $\cA$ by inner automorphisms.
\end{prop}

\proof
Consider a generator $\delta_r$, for $r\in \Q^*_+$, of the algebra $C^*_r(\Q^*_+)$.
The operator $e^{it H} \delta_r e^{-it H}$ acts on a basis element $\epsilon_{r'}$ as
$$ e^{it H} \delta_r e^{-it H} \epsilon_{r'} = n(rr')^{it} n(r')^{-it} \delta_r \epsilon_{r'}, $$
where for $r=p_1^{n_1}\cdots p_k^{n_k}$ in $\Q^*_+$, with $n_i\in \Z$, we have
$n(r)=p_1^{|n_1|}\cdots p_k^{|n_k|}$ in $\N$. If $r'=a/b$, with $a,b\in \N$ with $(a,b)=1$, 
and $r=u/v$ with $u,v\in \N$ with $(u,v)=1$, then 
$$ \frac{n(r'r)}{n(r')} = n(r) \cdot (b,u) \cdot (a,v). $$
Thus, for $r=u/v$, we can rewrite the operator above as
$$ e^{it H} \delta_r e^{-it H}  = \sum_{k | u} \sum_{\ell |v} n(r)^{it} k^{it} \ell^{it} \delta_r \Pi_{(k,\ell)}, $$
where $\Pi_{(k,\ell)} \epsilon_{r'}=\epsilon_{r'}$ if $k|a$ and $\ell |b$ and zero otherwise, where $r'=a/b$.
The term of the sum with $k=1$ and $\ell=1$ corresponds to the operator $n(r)^{it}\delta_r$.
Thus, this shows that the smallest $C^*$-subalgebra $\cA\subset \cB(\ell^2(\Q^*_+))$ 
that contains $C^*_r(\Q^*_+)$ and that is invariant under the  time evolution \eqref{sigmaBl2Q}
is the $C^*$-subalgebra $\cA\subset \cB(\ell^2(\Q^*_+))$ generated by the $\delta_r$ and by
the projections $\Pi_{(k,\ell)}$. Let $\Pi_n$ be the spectral projections of the operator $H$ corresponding
to the eigenvalues $\log(n)$ with $n\in\N$. We see that these are in the
algebra $\cA$ generated by the $\delta_r$ and the $\Pi_{(k,\ell)}$. Indeed
we have $\Pi_n \epsilon_r=\epsilon_r$ when $n=n(r)$ and zero otherwise,
so that we have $\Pi_n =\sum_{k,\ell\,:\, k\ell =n} \Pi_{(k,\ell)}$. The unitary operator $e^{itH}$
is a bounded operator that is in the $C^*$-algebra generated by the spectral projections of $H$.
Thus, the time evolution \eqref{sigmaBl2Q} acts on the algebra $\cA$ by inner automorphisms.
\endproof

\smallskip
\subsection{Semigroup and Grothendieck group}\label{semiKgroupGsec}

We will see later in the paper that, in addition to the abelian semigroup
$(\cK,\#)$ of oriented knots with the connected sum operation, we also
need to consider the associated Grothendieck group. 

\smallskip

Let $\cG_\cK$ denote the universal enveloping abelian group
(Grothendieck group) of the semigroup $(\cK,\#)$. The decomposition into prime knots shows
that $(\cK,\#)$ is a free abelian semigroup on a countable set of generators given by the
prime knots. Thus, $(\cK,\#)$ is non-canonically isomorphic to the semigroup $(\N,\cdot)$,
and its enveloping group $\cG_\cK$ is non-canonically isomorphic to the multiplicative group
$\Q^*_+$. The universal enveloping abelian group $\cG_\cK$ of $(\cK,\#)$ can be identified 
with pairs $(K,K')$, up to the equivalence relation $(K,K')\sim (K\# K'',K'\# K'')$ for all $K''\in \cK$.
We write the equivalence classes of pairs as formal differences, denoted by $K\ominus K'$.

\smallskip

In the case of the semigroup $\cK_a$ of alternating knots with the connected sum
operation, freely generated by the set $\cP_{\cK, a}$ of alternating prime knots, we similarly
construct the enveloping abelian group $\cG_{\cK,a}$. It is also non-canonically isomorphic
to $\Q^*_+$.

\smallskip
\subsection{Statistical physics of the group $\cG_{\cK,a}$}\label{statKaSec}

We show that the construction presented above of a quantum statistical
mechanical system for $\Q^*_+$ with partition function $\zeta^2(\beta)/\zeta(2\beta)$,
with $\zeta(\beta)$ the Riemann zeta function, can be generalized to the case of 
the group  $\cG_{\cK,a}$. 

\smallskip

Let $K\ominus K'=(a_1 K_1\#\cdots \# a_j K_j)\ominus 
(b_1 K_1'\# \cdots \# b_\ell K'_\ell)$ be an element of $\cG_{\cK,a}$ with primary
decompositions $K=a_1 K_1\#\cdots \# a_m K_m$ and $K'=b_1 K_1'\# \cdots \# b_\ell K'_\ell$,
where the $K_i$ and $K'_j$ are all distinct prime knots, with multiplicities $a_i$ and $b_j$.
Let $\epsilon_{K\ominus K'}$ be the corresponding basis element of $\ell^2(\cG_{\cK,a})$.
For a knot $K$, let $\omega(K)$ denote the number of distinct prime knots in its
primary decomposition, namely $\omega(K)=m$ for $K=a_1 K_1\#\cdots \# a_m K_m$
with the $K_i$ prime.

\begin{prop}\label{propHGKa}
Consider the operator $H$ acting on $\ell^2(\cG_{\cK,a})$, which acts on basis elements as
\begin{equation}\label{HGKa}
H \, \epsilon_{K\ominus K'} = \left(\sum_{i=1}^m (a_i (Cr(K_i)+g(K_i)) + \sum_{j=1}^\ell 
b_j (Cr(K'_j)+g(K'_j))) \right) \log(q) \, 
\epsilon_{K\ominus K'}.
\end{equation}
This is an unbounded densely defined operator, such that $e^{-\beta H}$ is trace class for
all $\beta \geq \beta_+$, satisfying
\begin{equation}\label{ZetaHGKa}
Z_{\cG_{\cK,a}}(\beta):=\Tr(e^{-\beta H}) = \frac{Z_a^2(\beta)}{Z_a(2\beta)},
\end{equation}
where $Z_a(\beta)$ is the partition function of Theorem \ref{ZetaCrg}.
\end{prop}

\proof The argument is very similar to the case of $\Q^*_+$ discussed above.
By Lemma \ref{EulProd}, we can write
$$ \frac{Z_a^2(\beta)}{Z_a(2\beta)} =\prod_{K\in \cP_{\cK,a}} \frac{1-q^{-2\beta (Cr(K)+g(K))}}
{(1-q^{-\beta (Cr(K)+g(K))})^2}. $$
We then write this as
$$ \prod_{K\in \cP_{\cK,a}} \frac{1+q^{-\beta (Cr(K)+g(K))}}{1-q^{-\beta (Cr(K)+g(K))}} = 
\prod_{K\in \cP_{\cK,a}} (\frac{1}{1-q^{-\beta (Cr(K)+g(K))}}  +\frac{q^{-\beta (Cr(K)+g(K))}}{1-q^{-\beta (Cr(K)+g(K))}}) $$ $$ =\prod_{K\in \cP_{\cK,a}} ( 1 + 2 \sum_{n\geq 1} q^{-\beta n (Cr(K)+g(K))} ). $$
On the other hand, we have
$$ \Tr(e^{-\beta H}) =\sum_{K\ominus K' \in \cG_{\cK,a}} \langle \epsilon_{K\ominus K'},
e^{-\beta H} \epsilon_{K\ominus K'}\rangle = \sum_{\lambda \in \Spec(H)} m_\lambda e^{-\beta \lambda}, $$
where $m_\lambda$ are the multiplicities. By \eqref{HGKa} the operator is diagonal on the
basis $\epsilon_{K\ominus K'}$ with eigenvalues $q^{-\beta ((Cr(K)+g(K))+(Cr(K')+g(K'))}$.
The multiplicities are $2^{m+\ell}$ for $K=a_1 K_1\#\cdots \# a_m K_m$ and 
$K'=b_1 K_1'\# \cdots \# b_\ell K'_\ell$, since all the other basis vectors in the same
eigenspace are obtained by moving some of the $K_i$ and $K_j'$ factors  to the other
side of $\ominus$, hence for each primary term in the decomposition there are two choices.
Thus, we obtain
$$ \Tr(e^{-\beta H}) =\sum_{K\in \cK_a} 2^{\omega(K)} \, q^{-\beta (Cr(K)+g(K))}. $$
We then see by rewriting this in the Euler product form that the identity \eqref{ZetaHGKa} holds.
\endproof

\smallskip

We have the analog of Proposition \ref{inneractQ}, which is proved by the same
argument.

\begin{prop}\label{inneractK}
The smallest $C^*$-subalgebra $\cA\subset \cB(\ell^2(\cG_{\cK,a}))$ 
that contains $C^*_r(\cG_{\cK,a})$ and is invariant under the time evolution 
\begin{equation}\label{sigmaBG}
\sigma_t(T) = e^{itH} T e^{-itH}, \ \ \  T\in \cB(\ell^2(\cG_{\cK,a})),
\end{equation}
with $H$ as in \eqref{HGKa}, 
is generated by $C^*_r(\cG_{\cK,a})$ and by projections 
$$ \Pi_{(K_1,K_2)} \epsilon_{K\ominus K'}=\left\{ \begin{array}{ll} 
\epsilon_{K\ominus K'} & K_1|K \text{ and } K_2 |K'\\
0 & \text{otherwise} \end{array}\right. $$
The time evolution \eqref{sigmaBG} acts on the algebra $\cA$ by inner automorphisms.
\end{prop}

The fact that this time evolution is inner is undesirable from the operator-algebraic
point of view. We will return to discuss this problem in the following sections. A first step
towards improving the system described in this section is to introduce interaction terms
in the quantum statistical mechanical system, as one does in the case of the Bost--Connes
system by passing from the algebra $C^*_r(\N)$ to the crossed product
algebra $C^*_r(\Q/\Z)\rtimes \N$. In our setting the analogous step will consist in
passing from a quantum statistical mechanical system associated to knots to one
associated to 3-manifolds.

\medskip
\section{Knot groups, $3$-manifolds, and cyclic branched covers}\label{3mfldSec}

\subsection{Cyclic branched coverings of the $3$-sphere}

Let $\cK$ denote the set of ambient isotopy classes of (oriented) knots in $S^3$. 
For simplicity of notation, in the following we will write $K$ for a knot and 
also for its equivalence class up to ambient isotopy. 

\smallskip

It is well known, \cite{Alex}, \cite{Hilden}, \cite{Monte}, 
that every smooth oriented closed $3$-manifold can be realized (non-uniquely) as a 
branched cover of the $3$-sphere, branched along a knot. Moreover, it is also
well known that an $n$-fold branched covering of the $3$-sphere, branched along
a knot $K$, is entirely determined by the datum of a representation 
\begin{equation}\label{reppi1KSn}
\rho: \pi_1(S^3\smallsetminus K) \to S_n,
\end{equation}
where $\pi_1(S^3\smallsetminus K)$ is the fundamental group of the
knot complement, and $S_n$ is the symmetric group of permutations
of $n$ elements, \cite{Alex}. 

\smallskip

An $n$-fold branched covering of the $3$-sphere $S^3$, branched along 
a knot $K$, is said to be {\em cyclic} or {\em abelian} if the corresponding homomorphism 
\eqref{reppi1KSn} factors through the abelianization 
$\pi_1(S^3\smallsetminus K)^{ab}=H_1(S^3\smallsetminus K, \Z)=\Z$, as a
homomorphism $\rho: H_1(S^3\smallsetminus K, \Z) \to \Z/n\Z$ with values in 
the subgroup of cyclic permutations $\Z/n\Z \subset S_n$. 
In particular, for a given knot $K$, there is a unique connected cyclic branched 
covering $Y_n(K)$. We write $\pi_{K,n}: Y_n(K)\to S^3$ for the corresponding
projection map. The remaining elements in 
$\Hom(H_1(S^3\smallsetminus K, \Z), \Z/n\Z)$ correspond to coverings that
have multiple components. 

\smallskip

It is known, \cite{BoilPaol}, \cite{Zimm},
that, if one knows the cyclic coverings $Y_p(K)$ for three
distinct primes $p$, this uniquely identifies the knot $K$. In other words, 
given a knot $K$, there are at most two distinct primes $p\neq p'$ for which there 
exist some inequivalent knot $K'$ with
homeomorphic branched cyclic coverings, $Y_p(K)\simeq Y_p(K')$ and
$Y_{p'}(K)\simeq Y_{p'}(K')$. 

\smallskip

To the purpose of building an analog of the Bost--Connes system in
the setting of arithmetic topology, 
we think of cyclic branched coverings of the 3-sphere $S^3$ as an
analog of abelian extensions of $\Q$.

\subsection{Knots semigroup}
For the purpose of our construction, we will consider a semigroup 
\begin{equation}\label{semigroupS}
\cS =\cK \times \N,
\end{equation}
where $\cK=(\cK,\#)$ is the semigroup of oriented knots with the direct
sum operation, as in the previous section, and $\N$ is the multiplicative semigroup of positive integers.
The idea behind this choice is that an element $(K,n)\in \cS$ specifies the
branch locus and of the order of a cyclic branched covering of $S^3$. 
The semigroup $\cS$ is generated by the pairs $(K,p)$ where $K$
is a prime knot and $p$ is a prime number. 

\smallskip

The quantum statistical mechanical model we discussed in the
previous section for the semigroup $\cK_a$ and its group
completion $\cG_{\cK,a}$ extend to the product $\cK_a\times \N$
as follows.

\begin{lem} \label{QSMKN}
Let $N: \cK_a\times \N \to \R^*_+$ be a semigroup homomorphism.
Then setting $\sigma_t(\mu_{K,n})=N(n,K)^{it} \mu_{K,n}$ defines a
time evolution of $C^*_r(\cK_a\times \N)$. In particular, taking
$$ N(K,n) = n \, q^{(Cr(K)+g(K))} $$
determines a time evolution with partition function
$\zeta(\beta) Z_a(\beta)$, where $\zeta(\beta)$ is the Riemann 
zeta function and $Z_a(\beta)$ is as in Theorem \ref{ZetaCrg},
for $\beta > \max\{ \beta_+(q), 1 \}$.
\end{lem}

\proof The argument is analogous to Theorem \ref{ZetaCrg}. The
time evolution $\sigma_t(\mu_{K,n})=N(n,K)^{it} \mu_{K,n}$ with
$N(K,n) = n \, q^{(Cr(K)+g(K))}$ is implemented by a Hamiltonian
of the form 
$$ H \epsilon_{K,n} = ( (Cr(K)+g(K)) \log (q) + \log(n)) \epsilon_{K,n} $$
on the canonical basis of $\ell^2(\cK_a\times \N)$, with partition
function
$$ Z_{\cK_a\times \N}(\beta) =\Tr( e^{-\beta H}) =\sum_{K,n} \langle \epsilon_{K,n},
e^{-\beta H} \epsilon_{K,n}\rangle = \sum_{K,n} q^{-\beta(Cr(K)+g(K))} n^{-\beta}
= Z_a(\beta) \cdot \zeta(\beta), $$
where $Z_a(\beta)$ is the partition function of Theorem \ref{ZetaCrg} and
$\zeta(\beta)$ is the Riemann zeta function. The operator $e^{-\beta H}$
is trace class in the range $\beta > \max\{ \beta_+(q), 1 \}$.
\endproof

\subsection{Wirtinger presentations and connected sums}

The fundamental group $\pi_1(S^3\smallsetminus K)$ of 
a knot complement has an explicit presentation, associated
to the choice of a planar diagram $D(K)$ representing the knot $K$.
It is given by the Wirtinger presentation $\cW(D(K))$. 
Let $N_D$ be the number of crossings
in the planar diagram $D=D(K)$. Then in the presentation 
$\cW(D(K))$ there are $N$ generators $a_i$, identified with 
loops circling around the oriented arcs given by the two parts of
the lower branch at each crossing (drawn as two arcs in the planar
diagram). At each crossing one imposes a relation, which is either
of the form $a_i a_j^{-1} a_{i+1}^{-1} a_j =1$ or $a_i a_j a_{i+1}^{-1} a_j^{-1}=1$,
depending on the orientations at the crossing, see \S 4.2.3 of \cite{Still}
for more details. The following fact is well known. We reproduce it here
for the reader's convenience.

\begin{lem}\label{Wirtconnsum}
Let $K=K_1\# K_2$ be a connected sum. Then the fundamental groups
satisfy
\begin{equation}\label{amalgpi1}
\pi_1(S^3\setminus(K_1\#K_2))=\pi_1(S^3\setminus K_1) *_{\Z}\pi_1(S^3\setminus K_2).
\end{equation}
\end{lem}

\proof Choose planar diagrams $D_1=D(K_1)$ and $D_2=D(K_2)$. Let $N_i$ 
be the number of crossings in $D_i$. In these
diagrams, let us number the arcs so that $a_1$ and $b_1$ are, respectively, the
arcs where the connected sum operation is performed. Let $D=D(K_1\# K_2)$ be the resulting planar
diagram for the connected sum knot. Let $\cW(D_1)=\langle a_1, \ldots, a_{N_1}\,|\,
r_1,\ldots, r_{N_1}\rangle$ and $\cW(D_2)=\langle b_1,\ldots, b_{N_2}\,|\,
s_1, \ldots, s_{N_2}\rangle$ be the Wirtinger presentations of 
$\pi_1(S^3\smallsetminus K_i)$ associated to these planar diagrams.  Let 
$f_i: \Z \to \pi_1(S^3\smallsetminus K_i)$ denote the homomorphisms that
map the generators of $\Z$ to the generators, in the respective Wirtinger presentations
as above, given by the arcs chosen for the connected sum: $f_1(1)=a_1$ and 
$f_2(1)=b_1$. The amalgamated product in \eqref{amalgpi1} is the resulting pushout diagram of groups 
\begin{equation}\label{pushdiagr} \begin{tikzcd}
\Z \arrow{r}{f_1} \arrow{d}{f_2}
&\pi_1(S^3\setminus K_1) \arrow{d}\\
\pi_1(S^3\setminus K_2) \arrow{r} & \pi_1(S^3\setminus K_1) *_{\Z}\pi_1(S^3\setminus K_2)
\end{tikzcd}
\end{equation}
where $\pi_1(S^3\setminus K_1) *_{\Z}\pi_1(S^3\setminus K_2)$ has a presentation
of the form $$\langle a_1,\cdots, a_{N_1}, b_1,\cdots, b_{N_2}\,|\, r_1,\cdots, r_n, s_1,\cdots, s_m,
a_1 b_1^{-1} \rangle,$$
which agrees with the Wirtinger presentation $\cW(D)$ of $D=D(K_1\# K_2)$,
hence the pushout group is isomorphic to $\pi_1(S^3\setminus(K_1\#K_2))$.
\endproof

\begin{cor}\label{cordirsys}
The groups $\pi_1(S^3\smallsetminus K)$ form a direct system, with respect to
the directed set $\cK$, partially ordered by divisibility with respect to the direct sum
operation, with maps
\begin{equation}\label{directsyspi1K}
\varphi_{K',K}: \pi_1(S^3\smallsetminus K')\to \pi_1(S^3\smallsetminus K),  \ \ \  \text{ for } K' | K,
\end{equation}
\end{cor}

\proof A knot $K'$ divides a knot $K$ in the
semigroup $(\cK,\#)$ if there is some other knot $K''$ such that
$K=K'\# K''$. Defining a partial order by setting $K'\leq K$ if $K'$ divides $K$
makes $\cK$ into a directed set.  As in the previous lemma, we then have 
a group homomorphism  $\varphi_{K',K}: \pi_1(S^3\smallsetminus K')\to \pi_1(S^3\smallsetminus K)$
given by the corresponding map in the pushout diagram \eqref{pushdiagr}. These
morphisms satisfy $\varphi_{K,K}=1$ and $\varphi_{K_2,K_3} \circ \varphi_{K_1,K_2}=\varphi_{K_1,K_3}$
when $K_1 | K_2$ and  $K_2 | K_3$, hence the groups $\pi_1(S^3\smallsetminus K)$  form 
a direct system. 
\endproof

We can then consider the direct limit of this direct system,
\begin{equation}\label{dirlimpi1}
\pi := \varinjlim_{K\in \cK} \pi_1(S^3\smallsetminus K) =\varinjlim_{K\in \cK} \pi_K, 
\end{equation}
where we use the shorthand notation $\pi_K:= \pi_1(S^3\smallsetminus K)$.
The direct limit $\pi$ is given by equivalence classes of elements $\gamma_K \in \pi_K$,
under the relation $\gamma_K \sim \gamma_{K'}$ if 
there is some $K''$ in $\cK$ such that $K|K''$ and $K' | K''$ with 
$\varphi_{K,K''}(\gamma_K) = \varphi_{K',K''}(\gamma_{K'})$.
Setting $[\gamma_K]\cdot [\gamma_{K'}]:= \varphi_{K, K\# K'}(\gamma_K)\cdot 
\varphi_{K', K\# K'}(\gamma_{K'})$, with the product in $\pi_{K\# K'}$ determines
a product on $\pi$ that is independent of representatives. 

\smallskip

In addition to considering the direct limit $\pi$ of the directed system of
the groups $\pi_K$ with the homomorphisms $\varphi_{K,K\# K'}$, it will
be convenient for our purposes to also consider the direct product
\begin{equation}\label{dirsumgroup}
\tilde \pi := \prod_{K\in \cK}\,\, \pi_1(S^3\smallsetminus K) =\prod_{K\in \cK}\,\, \pi_K,
\end{equation}
without imposing the equivalence relations of the direct limit. There are
then induced morphisms 
\begin{equation}\label{phiKtildepi}
\varphi_{K'}: \tilde\pi \to \tilde\pi, \ \ \ \  \varphi_{K'}=(\varphi_{K,K\# K'})_{K\in \cK}
\end{equation}
given coordinatewise by the morphisms $\gamma_K \mapsto \varphi_{K,K\# K'}(\gamma_K)$ 
of the direct system.

\smallskip
\subsection{Wild knots and fundamental groups}

Wilder knots are a class of wild knots with a single wild point,
obtained as infinite connected sums of a sequence $K_n$ of tame knots.
It is well known (see \cite{Crow}, \cite{Lomo}) that such Wilder knots
have knot group isomorphic to the infinite amalgamated product
$\pi_\infty:=\pi_1(S^3\smallsetminus K_\infty)=\pi_{K_1}*_\Z \pi_{K_2} *_\Z \pi_{K_3}\cdots
*_\Z \pi_{K_N}*_\Z \cdots$. More generally wild knots have knot groups that are
obtained as direct limits of knot groups of tame knots, \cite{Crow}.
The direct limit $\pi=\varinjlim_K \pi_K$ described above has a similar interpretation. 

\begin{lem}\label{piwild}
The direct limit $\pi=\varinjlim_K \pi_K$ is the knot group $\pi=\pi_1(S^3\smallsetminus K_\infty)$
of a wild knot $K_\infty$ with a Cantor set of wild points.
\end{lem}

\proof The construction of the wild knot $K_\infty$ is modeled on the direct system of
groups $\pi_K$ under the order relation in the semigroup $\cK$ given by divisibility.
Choose an enumeration of the prime knots. Any such choice determines a  bijection
between the set of prime knots and the set of prime numbers, and a corresponding
isomorphisms of semigroups $(\cK,\#)\simeq (\N,\cdot)$. We write the chosen
enumeration of the prime knots as $\{ K_p \}$ where $p$ ranges over prime numbers. 
Starting with the unknot in $S^3$, construct a Wilder knot given by the infinite connected 
sum of all the prime knots $K_p$. This has a single wild point lying on the initial unknot.
At the successive step repeat the procedure in each of the prime knots of the previous level,
namely insert in each the full sequence of prime knots $K_p$, with a single tame point for
each knot of the previous level, which we locate at the
intersection of those knots with the original unknot. In the limit the resulting wild knot $K_\infty$ has 
set of wild points that is compact, totally disconnected, with each wild point
an accumulation point of other wild points. The fundamental group of
the knot complement of $K_\infty$ is then obtained as in \cite{Crow} as the direct limit $\pi$.
\endproof

\smallskip

\begin{rem}\label{treeKinfty}{\rm
In a rooted tree, we say that a vertex has level $N$ if it is connected to
the root by a path of $N$ edges.
Let $\cT$ be the non-locally-finite labelled rooted tree with root vertex
labelled by the unknot. The root vertex (level zero) is connected to a countable
infinity of vertices labelled by the prime knots (level one). in turn each of these
vertices is connected to another countable set of vertices labelled by the prime
knots (level two), and so on, with each vertex at level $N$ connected to a
countable set of vertices at level $N+1$, labelled by the prime knots. 
The number of vertices at a given level $N$ is a countable union of countable
sets, hence countable. Let $E\cT$ be the resulting infinite set of edges.
The wild knot $K_\infty$ of Lemma \ref{piwild} can be described as
obtained by performing a connected sum along each of the edges of the tree
$\cT$.  }
\end{rem}

\smallskip
\subsection{Projective limits and cyclic coverings}\label{ProjLimSec}

For an arbitrary knot $K$, the abelianization of the fundamental group
$\pi_K= \pi_1(S^3\smallsetminus K)$ is always just the infinite cyclic
group generated by the meridian
\begin{equation}\label{H1Z}
\pi_1(S^3\smallsetminus K)^{ab}= H_1(S^3\smallsetminus K)=\Z.
\end{equation}
In particular, by the form of the relations in the Wirtinger presentation,
one sees that any representation of the group $\pi_K$ into an abelian
group $H$ will necessarily map all the generators of $\pi_K$ to a 
same element of $H$. 

\smallskip

Let us consider again the representation
$\rho=\rho_{K,n}: \pi_1(S^3\smallsetminus K) \to \Z/n\Z$ that
corresponds to the unique connected cyclic branched cover $Y_n(K)$
of $S^3$, branched along $K$. This representation sends all the
generators of $\pi_K$ to a primitive $n$-th root of unity, and it corresponds
to the quotient homomorphism $\rho_{K,n}: \Z=\pi_1(S^3\smallsetminus K)^{ab} \to \Z/n\Z$. 

\smallskip

Consider again the maps $\sigma_m: \Z/nm \Z \to \Z/n\Z$ that raise to the $m$-th power
and determine the projective system of the $\Z/n\Z$, with the indices $n\in \N$ ordered
by divisibility, with limit the profinite completion of $\Z$,
$$ \varprojlim_{n\in \N} \Z/n\Z =\hat\Z. $$
We have the simple compatibility condition of the maps $\rho_{K,n}$,
\begin{equation}\label{rhodiag}
\begin{tikzcd}
\Z=H_1(S^3\setminus K,\Z) \arrow{r}{\rho_{nm}} \arrow{rd}{\rho_n} & \Z/nm\Z \arrow{d}{\sigma_{m}} \\
& \Z/n\Z .
\end{tikzcd}
\end{equation}
The induced map to the projective limit is just the canonical map $\rho: \Z \to \hat\Z$ of the
integers to their pro-finite completion. 

\smallskip

Using the identification $\hat\Z=\Hom(\Q/\Z,\Q/\Z)$, we can think of the resulting map
$\rho: \Z=H_1(S^3\setminus K,\Z)\to \hat\Z$ as describing a locally trivial fibration over 
$S^3\smallsetminus K$ with fiber the set of all roots of unities, identified with $\Q/\Z$,
extended to $S^3$ with branch locus $K$. 
This space can be regarded as a limit, in the category of topological spaces,
of the cyclic branched coverings $Y_n(K)$. We denote it by $Y_{\hat\Z}(K)$.

\smallskip

When we consider simultaneously the inverse limit of the fibers, and direct limit of
the branch loci knots, we obtain a space $Y_{\hat\Z}(K_\infty)$, which is a branched
cover of $S^3$ branched along the wild knot $K_\infty$ with fiber the set of roots of
unity. The covering is specified by a representation of the direct limit group $\pi$ 
to $\hat\Z$. The space $Y_{\hat\Z}(K_\infty)$ is the geometric object underlying the
construction of the quantum statistical mechanical system that we describe in the
coming section.

\medskip
\section{Quantum Statistical Mechanics of $3$-manifolds} \label{3mfldQSMSec}

In this section we combine the constructions of the previous section with
the quantum statistical mechanics of knots, to construct an analog of the
Bost--Connes system associated to cyclic branched coverings of
the $3$-sphere. We show that the properties of the resulting quantum
statistical mechanical system are significantly different from the
Bost--Connes case and are related to noncommutative Bernoulli crossed
products.

\smallskip
\subsection{Group rings}

Thus, in order to construct a replacement for the algebra $C^*(\Q/\Z)=C(\hat\Z)$
of the Bost--Connes system, which will account for all the possible choices of a
knot $K$ and a cyclic branched cover of some order $n$, we need to introduce
appropriate group rings. We first deal with the part of the information that concerns
the knot complements and the knot groups $\pi_K=\pi_1(S^3\smallsetminus K)$
and then, in \S \ref{SecKhatZ} below, we combine this part of the construction
with the information on the choice of the cyclic branched coverings coming from the
$\hat \Z$ datum.

\smallskip

We consider group rings $\Q[\pi_K]$, for each knot group $\pi_K=\pi_1(S^3\smallsetminus K)$,
and also the group ring $\Q[\tilde \pi]$, with $\tilde\pi$ the direct product of the $\pi_K$
as in \eqref{dirsumgroup}, and the group ring $\Q[\pi]$, with $\pi$ the direct limit of the $\pi_K$.

\smallskip

Note that, unlike the group ring $\Q[\Q/\Z]$ of the Bost--Connes system, the
group rings $\Q[\pi_K]$, $\Q[\tilde\pi]$ and $\Q[\pi]$ are noncommutative, hence
the corresponding $C^*$-algebra completions, which we will discuss later, 
can no longer be written as algebras of continuous function on a dual group.
If one considers the abelianization $\pi^{ab}$, 
all the maps of the direct system of the groups $\pi_K$ induce the identity on
the homology groups, hence $\pi^{ab}=\Z$, and one would simply obtain
the commutative group ring $\Q[\pi^{ab}]=\Q[\Z]=\Q[t,t^{-1}]$.

\smallskip
\subsection{Semigroup action and crossed product}

In the following, we consider the semigroup $\cK$ acting
as endomorphisms of $\Q[\tilde \pi]$ via the morphisms
\begin{equation}\label{sigmaK}
 \sigma_K: \gamma_{K'} \mapsto \varphi_{K', K\#K'}(\gamma_{K'}), 
\end{equation} 
for $\gamma_{K'}$ in $\pi_{K'}\subset \tilde\pi$.

\smallskip

As we have seen in Corollary \ref{cordirsys}, the maps $\varphi_{K,K\#K'}$
satisfy $\varphi_{K\#K', K\# K'\# K''}\circ \varphi_{K,K\#K'}=\varphi_{K,K\# K'\# K''}$.
Thus, the homomorphism $\sigma_K: \tilde\pi \to \tilde\pi$ defined by \eqref{sigmaK}
is indeed a semigroup action, since we have
$$  \sigma_{K_1\#K_2}(\gamma_{K'})=\varphi_{K', K_1\#K_2\#K'}(\gamma_{K'}) $$
$$ =\varphi_{K_1\#K',K_1\#K_2\#K'}(\varphi_{K', K_1\#K'}(\gamma_{K'}))=
\sigma_{K_2}(\sigma_{K_1}(\gamma_{K'})) .$$
We use the same notation for the induced morphism of the group ring
$\sigma_K: \Q[\tilde\pi] \to \Q[\tilde\pi]$.

\smallskip

By direct inspection of the respective Wirtinger presentations, as in Lemma \ref{Wirtconnsum},
we see that the generators $\{ a, a_2, \ldots, a_{N_1}, b_2,\ldots, b_{N_2} \}$ of $\pi_{K_1\# K_2}$
satisfy $a=\varphi_{K_1,K_1\# K_2}(a_1)=\varphi_{K_2,K_1\# K_2}(b_1)$. The remaining generators
$a_i=\varphi_{K_1,K_1\# K_2}(a_i)$ have a preimage in $\pi_{K_1}$ but no preimage in $\pi_{K_2}$,
and vice versa for the $b_i$. Thus, an element $\gamma_{K_1\#K_2}\in \pi_{K_1\#K_2}$
has either one preimage or none in $\pi_{K_1}$ and $\pi_{K_2}$. Let $\cR_K$ denote
the range of the homomorphism $\sigma_K$ acting on $\Q[\tilde \pi]$. As a subring of $\Q[\tilde \pi]$,
$\cR_K$ is generated by all the elements $\gamma_{K\# K'}$ in some $\pi_{K\# K'}\subset \tilde\pi$
that are in the range of $\varphi_{K', K\#K'}$. Then, by the observation above, there is a ring homomorphism
$\eta_K: \cR_K \to \Q[\tilde\pi]$ given by $\eta_K(\gamma_{K\# K'})=\gamma_{K'}$ for
$\gamma_{K\# K'}=\varphi_{K', K\#K'}(\gamma_{K'})$, satisfying $\sigma_K\circ \eta_K=id|_{\cR_K}$
and $\eta_K \circ \sigma_K =id|_{\Q[\tilde\pi]}$. 

\smallskip

\begin{rem}\label{sigmainj}{\rm
The behavior of the endomorphisms $\sigma_K$ here is significantly different from the
case of the endomorphisms $\sigma_n$ of the Bost--Connes system. Indeed, the $\sigma_K$ are
injective, while the $\sigma_n$ are surjective. The case we are looking at here resembles
closely the adaptation of the Bost--Connes system to the Habiro ring considered in \cite{Mar},
where a similar injectivity condition is satisfied by the $\sigma_n$. Our construction here
follows closely the setting of \cite{Mar} and of \S 4.7 of \cite{MaTa}.}
\end{rem}

\smallskip

Let $\cG_\cK$ denote the universal enveloping abelian group
(Grothendieck group) of the semigroup $(\cK,\#)$, as in \S \ref{semiKgroupGsec}.

\smallskip

\begin{lem}\label{dirlimQpi}
The direct limit of the ring homomorphisms $\sigma_K : 
\Q[\tilde\pi] \to \Q[\tilde\pi]$ satisfies
\begin{equation}\label{dirlimrings}
\varinjlim_{K\in \cK} \left( \sigma_K : 
\Q[\tilde\pi] \to \Q[\tilde\pi] \right) \cong \bigotimes_{h\in \cG_\cK} \Q[\pi].
\end{equation}
\end{lem}

\proof First note that, since $\pi$ is the direct limit of the groups $\pi_K$, the
group ring $\Q[\pi]$ is the direct limit of the group rings $\Q[\pi_K]$.  Moreover,
because $\tilde\pi$ is the direct product of the groups $\pi_K$, the group ring is a tensor
product $\Q[\tilde\pi]=\otimes_{K\in \cK} \Q[\pi_K]$. Let $\psi_K: \pi_K \to \pi$ be the
maps to the direct limit determined by the direct system. They satisfy 
$\psi_{K\#K'}\circ\varphi_{K,K\# K'}=\psi_K$, for all $K,K'\in \cK$. We denote by the same symbol the resulting morphisms on the group rings. We have commutative diagrams
\begin{equation}\label{dirlimdiagr} \begin{tikzcd}
\otimes_K \Q[\pi_K] \arrow{r}{\psi=(\psi_K)} \arrow{d}{\sigma_{K'}}
& \otimes_K \Q[\pi] \arrow{d}{\hat\sigma_{K'}}\\
\otimes_K \Q[\pi_K] \arrow{r}{\psi=(\psi_K)} & \otimes_K \Q[\pi] ,
\end{tikzcd}
\end{equation}
where on the right hand side the morphism $\hat\sigma_{K'}: \otimes_K \Q[\pi] \to \otimes_K \Q[\pi]$
shifts the indices, mapping the copy of $\Q[\pi]$ in the $K$-th position to the copy in the
$K\#K'$-th position. Since the maps $\psi_K$ are the maps to the direct limit of the system of
the $\pi_K$, the direct limit of the system on the left column reduces to that of the right column,
or equivalently, the induced morphism between the direct limits is an isomorphism
$$ \psi: \varinjlim_{K'\in \cK} \left( \sigma_{K'} : 
\Q[\tilde\pi] \to \Q[\tilde\pi] \right) \stackrel{\cong }{\longrightarrow} \varinjlim_{K'\in \cK} \left(\hat\sigma_{K'}: 
\otimes_K \Q[\pi] \to \otimes_K \Q[\pi] \right). $$
Elements in the limit on the right hand side are rational combinations of 
equivalence classes of elements $g_{K,K'} \in \pi$ with
$K,K'\in \cK$ under the equivalence relation induced by the maps $\hat\sigma_{K''}$, given 
by the shifting of indices $g_{K,K'}\sim g_{K\#K'', K'\# K''}$, where 
$(K, K')\sim (K\# K'', K'\# K'')$ is the relation 
that defines the elements $h=K\ominus K'$ in the 
Grothendieck group $\cG_\cK$ of the abelian semigroup $\cK$. Thus, 
we can identify 
$$ \varinjlim_{K'\in \cK} \left(\hat\sigma_{K'}: 
\otimes_K \Q[\pi] \to \otimes_K \Q[\pi] \right) = \otimes_{h\in \cG_\cK} \Q[\pi]. $$
\endproof

\smallskip

We now consider a crossed product construction analogous to the version
of the Bost--Connes construction given in \cite{Mar}.

\begin{defn}\label{algcrossprod}
Let $\cA_{\tilde\pi,\cK}$ be the $\Q$-algebra generated by $\Q[\tilde\pi]$ and 
generators $\mu_K$, $\mu_K^*$
for $K\in \cK$ with the relations $\mu_K^* \mu_K=1$ and
\begin{equation}\label{relmuK}
\mu_K \sigma_K(\gamma_{K'}) = \gamma_{K'} \mu_K, \ \ \ \  \mu_K^* \gamma_{K'} 
=\sigma_K(\gamma_{K'}) \mu_K^*.
\end{equation}
\end{defn}

\smallskip

\begin{rem}\label{eK}{\rm
Unlike what happens with the Bost--Connes algebra, the elements $e_K=\mu_K \mu_K^*$
does not belong to the algebra $\Q[\tilde\pi]$. However, as we see below, these elements 
belong to the direct limit $\varinjlim_{K\in \cK} \left( \sigma_K : 
\Q[\tilde\pi] \to \Q[\tilde\pi] \right)$ described above. }
\end{rem}

\smallskip

Let $\cA_K$ be the rings generated by all the elements of the form $\mu_K \gamma_{K'} \mu_K^*$
with $\gamma_{K'}\in \tilde\pi$. When $K$ is the unknot we just have $\Q[\tilde\pi]$. A direct analog
of Lemma 2.2 of \cite{Mar} shows that the endomorphisms $\sigma_K: \Q[\tilde\pi] \to \Q[\tilde\pi]$
extend to morphisms $\sigma_K: \cA_{K'} \to \cA_{K'}$ when $K \nmid
K'$ and to morphism
$\sigma_K : \cA_{K'}\to \cA_{K''}$ when $K|K'$ with $K'=K\# K''$. Setting 
$\alpha_K(a)=\mu_K a \mu_K^*$ gives homomorphisms (since $\mu_K^*\mu_K=1$) 
mapping $\alpha_K: \cA_{K'} \to \cA_{K\# K'}$ satisfying
$$ \alpha_K(\sigma_K(a))=e_K a e_K, \ \ \  \sigma_K (\alpha_K(a)) =a, $$
where the idempotents $e_K=\mu_K \mu_K^*$ map $\Q[\tilde\pi]$ by $\gamma_{K'} \mapsto
e_K \gamma_{K'} e_K$ to the subring $e_K \cR_K e_K \subset \cA_K$, with $\cR_K$ the
range of $\sigma_K$ as above. All this can be seen easily by essentially the same argument 
as in Lemma 2.2 of \cite{Mar}. Moreover, as in Lemma 2.3 of \cite{Mar} we then have the
following identification.

\begin{lem}\label{algdirlim}
The algebra $\cA_{\tilde\pi,\cK}$ described above is the direct limit 
$\varinjlim_{K\in \cK} \left( \sigma_K : \Q[\tilde\pi] \to \Q[\tilde\pi] \right)$.
\end{lem}

\proof There are homomorphisms $\cA_K \hookrightarrow \cA_{K'}$ whenever $K'=K\# K''$ in $\cK$,
determined by identifying $\mu_K a \mu_K^* = \mu_{K'} \sigma_{K''}(a) \mu_{K'}^*$. Thus, we 
can identify the algebra $\cA_{\tilde\pi,\cK}$ with the direct limit
$$ \cA_{\tilde\pi,\cK} = \varinjlim_{K\in \cK} \cA_K =\cup_{K\in \cK}\, \cA_K. $$
The morphisms $\sigma_K$ and $\alpha_K$ described above are compatible with
the direct system of the $\cA_K$ and determine an invertible morphism between
the direct limits, hence giving the identification 
$$ \varinjlim_{K\in \cK} \cA_K \cong \varinjlim_{K\in \cK} \left( \sigma_K : \Q[\tilde\pi] 
\to \Q[\tilde\pi] \right). $$
\endproof

\smallskip

In particular, as in Lemma 2.3 of \cite{Mar} we see that the morphisms induced by
the $\sigma_K$ on the direct limit become invertible. 

\begin{lem}\label{sigmaKinv}
The maps induced on the direct limit $\otimes_{\cG_\cK}\Q[\pi]$ by the 
$\sigma_K:\Q[\tilde\pi]\to \Q[\tilde\pi]$ are isomorphisms.
\end{lem}

\proof In terms of the algebra $\cA_{\tilde\pi,\cK}$, the elements $e_K =\mu_K \mu_K^*$
are  idempotents, hence we can write them as $e_K =1 -p_K$ for some projection $p_K$.
Using the relations \eqref{relmuK} we see that these satisfy $\sigma_K(p_K)
=\mu_K^* (1-\mu_K \mu_K^*)\mu_K =0$. By the injectivity of the $\sigma_K$ this
gives $e_K=1$. Thus, the $\mu_K$ satisfy both $\mu_K^* \mu_K=1$ and $\mu_K \mu_K^*=1$
are therefore unitaries, not just isometries. The $\sigma_K$ are then automorphisms
with inverses $\alpha_K$. Equivalently, in terms of the direct system 
$\sigma_K:\Q[\tilde\pi]\to \Q[\tilde\pi]$, elements in the direct limit are sequences 
$g_{K\ominus K'}$, with formal differences $K\ominus K'\in \cG_\cK$, where 
$g_{K\ominus K'\# K''}=\hat\sigma_{K''}(g_{K\ominus K'})$, 
hence in the direct limit the maps induced by the $\sigma_K$ are surjective as well as injective. 
\endproof

\smallskip

Thus, the resulting crossed product algebra is a group crossed product, which is
just given by the Bernoulli action that shifts the tensor factors indices,
\begin{equation}\label{crossQalg}
\bigotimes_{h\in \cG_{\cK}} \Q[\pi] \, \rtimes \cG_{\cK}. 
\end{equation}

\smallskip
\subsection{Operator algebras: von Neumann algebra}

Given a discrete group $\Gamma$, one can consider the
action by bounded operators on the Hilbert space $\ell^2(\Gamma)$ given by the 
left (or right) action of the group on itself. This determines a representation
of the algebra $\C[\Gamma]=\Q[\Gamma]\otimes_\Q \C$ on $\ell^2(\Gamma)$. We
drop the explicit labeling of the left/right regular representation, and simply write
$R: \C[\Gamma] \to \cB(\ell^2(\Gamma))$. The reduced group $C^*$-algebra
$C_r^*(\Gamma)$ is the norm completion of $R(\C[\Gamma])$ in $\cB(\ell^2(\Gamma))$
and the von Neumann algebra $\cN(\Gamma)$ is the double commutant 
$R(\C[\Gamma])''$. The group von Neumann algebra $\cN(\Gamma)$ has a finite trace
given by $\tau(R(\gamma))=1$ if $\gamma=1$ and $\tau(R(\gamma))=0$ otherwise. 
Every von Neumann algebra can be decomposed as a direct integral of factors.
A group von Neumann algebra $\cN(\Gamma)$ is a factor if and only if $\Gamma$
has the infinite conjugacy classes (ICC) property, namely the conjugacy classes of all
nontrivial elements $\gamma\neq 1$ in $\Gamma$ are infinite. 

\smallskip

The question of whether the knot groups $\pi_K$ (in the non-torus case) satisfy the ICC
property was stated as an open problem (Problem 3) in \cite{delaHarpe}. It was then proved
in Corollary 11.1 of \cite{HarpePreaux} that indeed the knot groups $\pi_K$ are ICC if and
only if the knot $K$ is not a torus knot. 
A direct product of groups is ICC if and only if each of its factors is, hence
the group $\tilde\pi$ is not ICC because the factors $\pi_K$ corresponding
to torus knots are not ICC. 

\smallskip

\begin{lem}\label{ICCpi}
The countably generated group $\pi=\varinjlim_K \pi_K$
has the ICC property.
\end{lem}

\proof First observe that the groups $\pi_K$, for any non-prime knot
$K$, have the ICC property. This follows immediately from the topological
property that all torus knots are prime knots, hence by the characterization
of Corollary 11.1 of \cite{HarpePreaux} the knot group of every non-prime knot
is ICC. Moreover, by Proposition 5.1 of \cite{HarpePreaux}, 
if $\Gamma=\Gamma_1*_{\Gamma_0} \Gamma_2$ is an
amalgamated product of discrete groups, with respect to a
common subgroup $\Gamma_0$ that is not of index 2 
(non-degenerate case), then $\Gamma$ has the ICC property
if at least one of the two groups $\Gamma_1$, $\Gamma_2$ is ICC.
(For a more general characterization of the ICC property for
amalgamated products see \S 5.6 of \cite{Preaux}.) 
As in Lemma \ref{piwild}, we identify the direct limit $\pi$ with the
knot group $\pi_1(S^3\smallsetminus K_\infty)$ of a wild knot $K_\infty$ 
obtained from a tree of connected sums. As we have seen in Lemma \ref{piwild},
after choosing an enumeration $K_p$ of the prime knots, 
we can describe $K_\infty$ as the result of constructing a necklace 
given by the infinite connected sum $K_{p_1}\# K_{p_2} \#\cdots \# K_{p_n} \#\cdots$
of the prime knots in the chosen order, followed iteratively by repeatedly
inserting by connected sum similar necklaces into each of the knots at the
previous stage, see Remark \ref{treeKinfty}. 
Consider a finite subset $K_{p_i}$, $i=1,\ldots, N$ of prime knots and
the tame knot obtained as their direct sum $K=K_{p_1}\#\cdots \# K_{p_N}$.
Let $\hat K_i$, $i=1,\ldots, N$ be the wild knots consisting of all the successive
iterative level to be inserted by connected sum into each of the $K_{p_i}$, 
and let $\hat K$ be the remaining wild knot given by the infinite connected sum
of the remaining prime knots $\hat K=K_{p_{N+1}}\# K_{p_{N+2}}\# \cdots$ and
all the successive iterative levels inserted into these. Then we can
describe the resulting wild knot as $K_\infty = K  \# \hat K\#\hat K_1\# \cdots \# \hat K_N$.
By the previous observations $\pi_K$ has the ICC property hence
the amalgamated product $\pi_{K \# \hat K}=\pi_K *_\Z  \pi_{\hat K}$ also does,
and the same applies to the remaining connected sums with the $\hat K_i$.
\endproof

\smallskip

\begin{rem}\label{amen}{\rm
As mentioned above, the ICC property for a group $\Gamma$ corresponds to
$\cN(\Gamma)$ being a $II_1$ factor. Moreover, it is known \cite{Co} that if
the group $\Gamma$ is amenable then $\cN(\Gamma)$ is isomorphic to the
hyperfinite type $II_1$ factor $R$. However,
knot groups are non-amenable (see \cite{Fuji}), even though they are $K$-amenable
(Theorem 5.18 of \cite{Mislin}).}
\end{rem}

\smallskip

After changing to $\C$-coefficients, the crossed product \eqref{crossQalg} also
has a von Neumann algebra completion
\begin{equation}\label{Bernoullicross}
\bigotimes_{h\in \cG_\cK} \, \cN(\pi) \, \rtimes \cG_\cK ,
\end{equation}
which is a special case of the class of Bernoulli crossed products
first studied in \cite{Co}, and more recently in \cite{PopaVaes}, \cite{VaVe}.
For simplicity of notation, here we just write $\otimes$ instead of the
commonly used $\overline\otimes$, for tensor products in
the von Neumann algebra context. 
The algebra \eqref{Bernoullicross} is represented on the Hilbert space
$L^2(\cN(\pi),\tau) \otimes \ell^2(\cG_\cK)$ or equivalently
$\ell^2(\pi) \otimes \ell^2(\cG_\cK)$.
Other representations can be constructed using a unitary representation $\cV$ of $\pi$
and replacing $\ell^2(\pi)$ with $\ell^2(\pi)\otimes \cV$. An explicit example of how
this twisting by a representation $\cV$ can be obtained is discussed briefly in the
following \S \ref{dRsec}.

\smallskip
\subsection{Twisting by de Rham representations}\label{dRsec}

Because the abelianizations 
of the knot groups are all equal to $\pi_K^{ab}=\Z$, one-dimensional
representations of $\pi_K$ by unitary operators correspond
to character homomorphisms $$\Hom(\pi_K, U(1))=\Hom(\Z,U(1))=U(1).$$
For each $K\in \cK$, consider then a choice of a phase $\theta_K\in \R/\Z$.
With the above identification, this determines a homomorphism, which we
still denote by $\theta_K: \pi_K \to U(1)$, which sends all the generators of 
$\pi_K$ to the same element $\lambda_K=\exp(2\pi i \theta_K)\in U(1)$. 

\smallskip

While the $1$-dimensional representations of $\pi_K$ are only of this
trivial nature, with all generators acting as the same phase factor $\lambda_K$,
it is well known that the knot groups $\pi_K$ have interesting higher
dimensional representations. In particular, already in the $2$-dimensional
case, one has an interesting family of representations, the so called
de Rham representations. In general, representations of $\pi_K$ are
related to roots of the Alexander polynomial. 

\smallskip

The de Rham representations of knot groups are homomorphisms
$\pi_K \to \GL_2(\C)$. For each root $r_K$ of the Alexander polynomial
$\Delta_K(t)$ of the knot $K$, there are, up to conjugation, $2k_r$ de Rham 
representations of $\pi_K$, where $k_r$ is the largest $k$ such that 
the $k$-th order Alexander polynomial (that is, the greatest common 
divisor of the determinants of the $(n - k + 1) \times (n - k + 1)$ minors 
of the Alexander matrix) satisfies $\Delta_k(r)\neq 0$, see \cite{deRham}.
In a de Rham representation associated to a root $r$ of the Alexander polynomial 
$\Delta_K(t)$ the generators of $\pi_K$ are represented as $2\times 2$-matrices of
the form
$$ \left( \begin{array}{cc} \sqrt r & x \\ 0 & \frac{1}{\sqrt r} \end{array}\right). $$
In order to avoid the abelian representations where $x$ is the same
for all generators, one only considers {\em based} representations,
where one of the $x$, say for the first generator in a given Wirtinger
presentation of $K$, is equal to zero, while all the others are nonzero. 
the list of the elements $x$ associated to 
a set of the remaining generators of $\pi_K$ gives a vector in the kernel of the 
Alexander matrix $A_K(t)$ at $t=r$, see \cite{Burde}, \cite{deRham}.

\smallskip

For a knot $K$ let $\cV_K$ be the representation of $\pi_K$ given by 
the complex vector space $\cV_K=\oplus_r \cV_{K,r}$, where $r$ 
ranges over roots of the Alexander polynomial $\Delta_K(t)$ and 
$\cV_{K,r}$ is a $2$-dimensional de Rham representation of $\pi_K$, 
constructed as above. We denote by $R_K=\oplus_r R_{K,r}$ the resulting 
representation of $\pi_K$ on the vector space $\cV_K$.

\begin{lem}\label{interPhilem}
For a connected sum $K=K_1\# K_2$, the
representation satisfies $\cV_{K_1\#K_2}=\cV_{K_1} \oplus \cV_{K_2}$
with $R_{K_1\#K_2}=R_{K_1} \oplus R_{K_2}$. Let $\Phi_{K_i,K_1\#K_2}$,
$i=1,2$, denote the inclusions of the direct 
factors $\cV_{K_i}$ in $\cV_{K_1\#K_2}$. Under the direct 
system of homomorphisms $\varphi_{K_i,K_1\#K_2}: \pi_{K_i} \to \pi_{K_1\# K_2}$,
the representations satisfy the compatibility condition
\begin{equation}\label{RKrphi}
\Phi_{K_i,K_1\#K_2} \circ R_{K_i}(\gamma_{K_i})  
=  R_{K_1\# K_2}(\varphi_{K_i,K_1\#K_2}(\gamma_{K_i})) \circ \Phi_{K_1,K_1\#K_2}  
\end{equation}
\end{lem}

\proof 
The Alexander polynomial is given by the determinant
$\Delta_K(t)=\det(V_K - t V_K^\tau)$, where $V_K$ is
a Seifert matrix of the knot. For a connected sum $K=K_1\# K_2$,
given Seifert matrices $V_{K_1}$ and $V_{K_2}$, the direct sum
$V_K=V_{K_1}\oplus V_{K_2}$ is a Seifert matrix for $K$. 
Correspondingly, the Alexander matrix $A_K(t)= V_K -t V_K^\tau$
also satisfies $A_K(t) =A_{K_1}(t)\oplus A_{K_2}(t)$ for $K=K_1\# K_2$.
Thus, the set of roots of $\Delta_{K_1\# K_2}(t)$ is the union of the 
sets of roots of $\Delta_{K_i}$, which implies that the vector space is
a direct sum $\cV_K=\cV_{K_1}\oplus \cV_{K_2}$. The vectors in
the kernel of the Alexander matrix at a given root also correspond to
those of $A_{K_1}(t)$ and $A_{K_2}(t)$, depending on the root, hence
the representations also split as $R_K = R_{K_1}\oplus R_{K_2}$.
To check the compatibility conditions, notice that we are working with
based representations. We can always assume that the generator in
the given Wirtinger presentation of $\pi_{K_1}$ with $x=0$ in the de Rham representation
corresponds to the arc where the connected sum is performed, which
matches then with the action of the corresponding generator of $\pi_{K_2}$,
so that the resulting representation given by $R_{K_1} \oplus R_{K_2}$
on $\cV_K$ is indeed a representation of the amalgamated product
$\pi_K=\pi_{K_1} *_\Z \pi_{K_2}$.
\endproof

\smallskip

\begin{rem}\label{unitaryandhigher} {\rm
If we want the elements of $\pi_K$ to be represented as unitary, rather than
just as invertible operators, then we should consider $SU(2)$ representations
of the knot group $\pi_K$, as in \cite{Klassen}, rather than $\GL_2(\C)$ 
representations as above. For our purposes two-dimensional representations
will suffice, but the construction we obtain can be generalized to the higher
dimensional representations obtained as in \cite{BoFri}, \cite{Jebali}, \cite{SilWil}.}
\end{rem}

\smallskip

\begin{cor}\label{dirlimdeRham}
The compatibility condition \eqref{RKrphi} satisfied by the de Rham representations $\cV_K$
implies that they induce a representation of the direct limit $\pi=\varinjlim_K \pi_K$ on the
space $\cV=\varinjlim_K \cV_K$, obtained as the direct limit under the direct sytem of
morphisms $\Phi_{K,K\#K'}$.
\end{cor}

\proof An element in $\cV$ is an equivalence class of elements $v_K\in \cV_K$ under
the relation $v_K \sim \Phi_{K,K\# K'}(v_K)$. Defining the action $[\gamma_K]\in \pi$ on
$[v_K]\in \cV$ as $R([\gamma_K]) [v_K]:= [ R_K(\gamma_K) v_K ]$ is well defined, since
the compatibility condition \eqref{RKrphi} implies that 
$$ \Phi_{K,K\#K'}( R_{K}(\gamma_{K})  v_K) 
=  R_{K\# K'}(\varphi_{K,K\#K'}(\gamma_{K}))  \Phi_{K,K\#K'} v_K . $$
\endproof

\smallskip
\subsection{Operator algebras: $C^*$-algebra}

We now consider the reduced $C^*$-algebras of the knot groups $\pi_K$ and
of the direct limit $\pi$. 

\begin{lem}\label{Cstarrpi}
The reduced group $C^*$ algebra of the direct limit $\pi=\varinjlim_K \pi_K$
satisfies
\begin{equation}\label{starprodCstar}
C^*_r(\pi) = \varinjlim_{K\in \cK} C^*_r(\pi_K) =\star_{C_r^*(\Z),\cT} C_r^*(\pi_K),
\end{equation}
where $\star_{C_r^*(\Z),\cT} C_r^*(\pi_K)$ denotes the infinite amalgamated
product of the reduced $C^*$-algebras $C_r^*(\pi_K)$ along the common
subalgebra $C_r^*(\Z)=C(S^1)$, performed as in the amalgamated
products of groups, along the tree $\cT$, in Remark \ref{treeKinfty}.
\end{lem}

\proof
By Proposition 2.5 of \cite{See}, the reduced $C^*$-algebra of the direct limit 
$\pi$ is an amalgamated product of $C^*$-algebras. More precisely, by 
Lemma \ref{piwild}, we identify the direct limit $\pi$ with the
knot group $\pi_1(S^3\smallsetminus K_\infty)$ of the wild knot $K_\infty$ 
obtained from a tree of connected sums obtained by successively inserting
with connected sums in each of the knots of a necklace given by the infinite 
connected sum of the prime knots additional necklaces of the same kind,
and so on iteratively, see Remark \ref{treeKinfty}. 
Thus, the direct limit group can be identified as an
infinite sequence of amalgamated products $\pi_K *_\Z \pi_{K'}$, over a common
subgroup $\Z$, corresponding to each successive connected sum $K\# K'$.
As shown in Proposition 2.5 of \cite{See}, the reduced $C^*$-algebra of a
countably infinite amalgamated product of discrete countable groups, all 
performed along a same common subgroup, is an amalgamated product
of $C^*$-algebras, $C^*_r(\pi) = \star_{C_r^*(\Z),\cT} C_r^*(\pi_K)$,
where the amalgamated products are performed in the same way
as for the groups, using the notation $\pi = *_{\Z,\cT} \pi_K$ to indicate the
infinite amalgamated product as in Lemma \ref{piwild} and
Remark \ref{treeKinfty}, with the connected sums performed
along the edges of the tree $\cT$ as in Remark \ref{treeKinfty}.
The reduced amalgamated free product of reduced group $C^*$-algebras
is taken with respect to the conditional expectations. Namely, by Theorem 2.2
of \cite{See}, given a family of unital $C^*$-algebras $A_j$ all containing a 
sub-$C^*$-algebra $B$ with $1\in B$. If there are conditional expectations 
$E_j: A_j \to B$ with faithful GNS representations, then there is a unique $C^*$-algebra
$A$, the amalgamated product of the $A_j$ along $B$, with the properties that
$B\subset A$ with $1_A\in B$, with a conditional expectation $E: A\to B$
with a faithful GNS representation; with inclusions $A_j \subset A$ extending
the inclusion $B\subset A$, so that $A$ is generated as a $C^*$-algebra
by the $A_j$, which form a free family of subalgebras, with the expectations 
given by restrictions $E|_{A_j}=E_j$. The freeness condition means that
$E(a_1\cdots a_n)=0$ whenever $a_i\in A_{j_i}$ with $j_i\neq j_{i+1}$
and all $a_i \in Ker(E)$. It is shown in Theorem 2.3 and Proposition 2.5 of
\cite{See} that these conditions hold in the case of amalgamated 
products of reduced group $C^*$-algebras as above. Lemma 2.6 of \cite{See}, together
with Lemma \ref{piwild} above, also shows that 
$C^*_r(\pi) = \varinjlim_{K\in \cK} C^*_r(\pi_K)$.
\endproof

\smallskip

At the level of $C^*$-algebras, one can similarly consider the crossed product
\begin{equation}\label{Bernoullicross}
\bigotimes_{h\in \cG_\cK} \, C^*_r(\pi) \, \rtimes \cG_\cK ,
\end{equation}
acting on the same Hilbert space $\ell^2(\pi)\otimes \ell^2(\cG_\cK)$. 
As in the case of von Neumann algebras above, 
we simply write $\otimes$ for the completed tensor products in the 
operator algebra context.

\smallskip
\subsection{The combined system}\label{SecKhatZ}

We now combine the previous construction, based on the direct system
of the knot groups $\pi_K$ and the action of the semigroup $\cK$, with the
information on the choice of the cyclic branched cover, by combining the
algebra constructed above with the Bost--Connes algebra, via the
representations $\rho_{K,n}: \pi_K \to \Z/n\Z$ that specify the 
unique connected cyclic branched cover $Y_n(K)$ of $S^3$ 
of order $n$, branched along $K$.

\smallskip

Let $\mmu_\infty$ be the group of all roots of unity of arbitrary order, which
we identify with $\mmu_\infty \simeq \Q/\Z$. For any $n$, let $\mmu_n \simeq \Z/n\Z$
be the group of roots of unity of order $n$, with $\mmu_n \subset \mmu_\infty$.

\begin{rem}\label{munnotation}{\rm
Notational warning: we avoid the more standard notation 
$\mu_n$ and $\mu_\infty$ for the groups of roots of unity, to avoid a conflict 
with the Bost--Connes notation, that we follow below, 
where $\mu_n$ is used for the isometries in the crossed
product algebra.}
\end{rem}

Any group homomorphism $\rho_K : \pi_K \to \mmu_\infty$ or
$\rho: \pi \to \mmu_\infty$ factors through the abelianizations
$\pi_K^{ab}=\Z$ and $\pi^{ab}=\Z$, hence it maps all the
generators to an element $\zeta\in \mmu_\infty$, of some order $n$.
Thus, the homomorphisms $\rho_K$ and $\rho$ determine representations  
$\rho_{K,n}:\pi_K \to \Z/n\Z$ and $\rho_n: \pi\to \Z/n\Z$.
Let $\cR\subset \Hom(\pi, \mmu_\infty)$ and $\cR_K\subset \Hom(\pi_K, \mmu_\infty)$ 
be the subsets of homomorphisms such that the corresponding $\rho_{K,n}$
and $\rho_n$ determine the unique connected cyclic branched cover.

\smallskip

Consider then the pullback diagrams of groups
\begin{equation}\label{pullpiKmuinfty} \begin{tikzcd}
\hat\pi_{K,n} \arrow{r} \arrow{d}
& \mmu_\infty \arrow{d}{\sigma_n} \\
\pi_K \arrow{r}{\rho_K} & \mmu_\infty
\end{tikzcd}
\end{equation}
and
\begin{equation}\label{pullpimuinfty} \begin{tikzcd}
\hat\pi_n \arrow{r} \arrow{d}
& \mmu_\infty \arrow{d}{\sigma_n} \\
\pi \arrow{r}{\rho} & \mmu_\infty
\end{tikzcd}
\end{equation}
where $\sigma_n :\mmu_\infty \to \mmu_\infty$ is the endomorphism
$\sigma_n: \zeta \mapsto \zeta^n$, that is, the homomorphism
$\sigma_n : \Q/\Z \to \Q/\Z$ mapping $\sigma_n: r \mapsto n r$.
The pullback groups are given by $\hat\pi_{K,n}=\{ (\gamma,\zeta)\in \pi_K \times \mmu_\infty\,|\, \rho(\gamma)=\zeta^n \}$ and 
$\hat\pi_n=\{ (\gamma,\zeta)\in \pi \times \mmu_\infty\,|\, \rho(\gamma)=\zeta^n \}$.

\smallskip

\begin{lem}\label{projlimhatpiKn}
Let $\cS \subset \N$ be a subsemigroup, with the partial ordering defined by the divisibility
relation. The groups $\hat\pi_{K,n}$ and $\hat\pi_n$ form projective systems with respect to $n\in \cS$,
with epimorphisms $\hat\sigma_{n/m}:
\hat\pi_{K,n}\to \hat\pi_{K,m}$ for $m|n$ in $\cS$, and similarly for the $\hat\pi_n$, 
with respective projective limits $\hat\pi_{K,\rho_K,\cS}$ and
$\hat\pi_{\rho,\cS}$, which depend both on the initial choice of the morphism
$\rho_K \in \cR_K$ (respectively, $\rho\in \cR$) and on the semigroup $\cS$. 
\end{lem}

\proof We illustrate the argument for $\pi_K$; the case of the direct limit $\pi$
is analogous. When $m|n$ in $\cS$ we have a commutative diagram
\begin{equation}\label{pullprojsys} \begin{tikzcd}
\hat\pi_{K,n} \arrow{rr}\arrow{ddr}\arrow{dr}{\hat\sigma_{n/m}} & & \mmu_\infty \arrow{d}{\sigma_{n/m}} \\
& \hat\pi_{K,m} \arrow{r} \arrow{d}
& \mmu_\infty \arrow{d}{\sigma_m} \\
& \pi_K \arrow{r}{\rho_K} & \mmu_\infty
\end{tikzcd}
\end{equation}
where the arrow $\hat\sigma_{n/m}: \hat\pi_{K,n}\to \hat\pi_{K,m}$ is determined by the
universal property. We have $\hat\sigma_{n/m}(\gamma,\zeta)=(\gamma, \sigma_{n/m}(\zeta))$,
with $\rho(\gamma)=\zeta^n=(\sigma_{n/m}(\zeta))^m$, hence we obtain a projective
system of epimorphisms $\hat\sigma_{n/m}: \hat\pi_{K,n}\twoheadrightarrow \hat\pi_{K,m}$
for $m|n$. The construction of these pullback diagrams and the groups $\hat\pi_{K,n}$ of the
projective system depend on the initial choice of the homomorphism 
$\rho_K \in \cR_K\subset \Hom(\pi_K,\mmu_\infty)$ and on the semigroup $\cS$, hence the resulting 
projective limit $\hat\pi_{K,\rho_K,\cS}=\varprojlim_{n\in \cS} \hat\pi_{K,n}$ also 
depends on $\rho_K$ and $\cS$.
\endproof

\smallskip

As mentioned above, the representation and $\rho_K\in \cR_K$ maps
the generators of $\pi_K$ to a single element $\zeta$ in 
the set $\cP(n_{\rho_K})$ of primitive roots of unity of some order $n_{\rho_K}$.
An arbitrary element $\gamma \in \pi_K$ maps to some 
$\rho(\gamma)=\zeta_{n_{\rho_K}}^{n_\gamma}\in \mmu_{n_{\rho_K}}\subset \mmu_\infty$.
Similarly for $\rho \in \cR$. 

\begin{defn}\label{Nrho}
Given $\rho\in \cR$ (respectively, $\rho_K\in \cR_K$), 
Let $\N_\rho\subset \N$ (respectively, $\N_{\rho_K}\subset \N$)
be the subsemigroup of $n\in \N$ with $(n,n_\rho)=1$ (respectively, $(n,n_{\rho_K})=1$),
that is, the multiplicative semigroup generated by those primes $p\in \cP$
that do not occur in the primary decomposition of $n_\rho$ (respectively, $n_{\rho_K}$).  
We use the notation $\hat\pi_\rho:=\hat\pi_{\rho, \N_\rho}$ and
$\hat\pi_{K,\rho_K}:=\hat\pi_{K,\rho_K,\N_{\rho_K}}$ 
for the corresponding projective limits.
\end{defn}

\smallskip

\begin{rem}\label{nthrootsgen}{\rm
The effect of passing to the pullbacks $\hat\pi_{K,n}$ and $\hat\pi_n$ 
is to introduce $n$-th roots for the elements 
of the knot groups $\pi_K$ and of their limit $\pi$. Indeed, for each element $\gamma$
of $\pi_K$, with $\rho(\gamma)=\zeta_{n_{\rho_K}}^{n_\gamma}$,  
there are $n$ corresponding elements in $\hat\pi_{K,n}$ of the form $(\gamma, \zeta)$ 
with $\zeta^n=\zeta_{n_{\rho_K}}^{n_\gamma}$.
The projective limits $\hat\pi_{K,\rho_K}$ and $\hat\pi_\rho$ contain roots of the elements of $\pi_K$
(or of $\pi$) for arbitrary order in $\N_{\rho_K}$ (respectively, $\N_\rho$).}
\end{rem}

\smallskip

\begin{rem}\label{Galrem}{\rm
The construction of the pullbacks $\hat\pi_{K,n}$ and $\hat\pi_n$ and 
projective limits $\hat\pi_{K,\rho_K}$ and $\hat\pi_\rho$ is analogous to the construction
of formal roots of Tate motives in \S 4.2 of \cite{LoMa}.
}\end{rem}

\smallskip

\begin{prop}\label{sigmanhatpi}
For all $n\in \N_\rho$, there are homomorphisms $\sigma_n: \hat\pi_\rho \to \hat\pi_\rho$
given by 
\begin{equation}\label{sigmanhatpi}
\sigma_n(\gamma,\zeta):=(\gamma, \zeta^n) .
\end{equation}
The maps $\{ \sigma_n \}_{n\in \N_\rho}$ of \eqref{sigmanhatpi}
determine an action of the semigroup $\N_\rho$ by endomorphisms of the group ring
$\Q[\hat\pi_\rho]$. The endomorphisms $\sigma_n$ have partial inverses
$\alpha_n: \Q[\hat\pi_\rho] \to \Q[\hat\pi_\rho]$,
\begin{equation}\label{alphanhatpi}
\alpha_n (\delta_{(\gamma,\zeta)}) = \frac{1}{n} \sum_{\eta\,:\, \eta^n =\zeta} \delta_{(\gamma,\eta)}
\end{equation}
satisfying $\sigma_n \circ \alpha_n (\delta_{(\gamma,\zeta)}) =\delta_{(\gamma,\zeta)}$
and $\alpha_n \circ \sigma_n (\delta_{(\gamma,\zeta)}) = e_n \cdot \delta_{(\gamma,\zeta)}$,
where $e_n = n^{-1} \sum_{\xi\,:\, \xi^n=1} \delta_{(1,\xi)}$ is an idempotent in $\Q[\hat\pi_\rho]$.
The case of $\hat\pi_{K,\rho_K}$ is analogous.
\end{prop}

\proof
An element $(\gamma, \zeta)$ belongs to $\hat\pi_n$ when 
$\rho(\gamma)=\zeta^n$, that is, $\zeta^n=\zeta_{n_\rho}^{n_\gamma}$.
An element $(\gamma, \zeta)$ with $\gamma \in \pi$ and $\zeta\in \mmu_\infty$ 
is in $\hat\pi_\rho$ when there is some $m\in \N_\rho$ such
that $\zeta^m =\zeta_{n_\rho}^{n_\gamma}$. That is, 
$\zeta \in \cup_{m\in \N_\rho} \sigma_m^{-1}(\mmu_{n_\rho})$. 
Suppose given $(\gamma, \zeta)\in \hat\pi_\rho$ and $n\in \N_\rho$. 
We need to check that the element $\sigma_n(\gamma,\zeta):=(\gamma, \zeta^n)$
is also in $\hat\pi_\rho$.
Let $m\in \N_\rho$ be such that $\zeta^m=\rho(\gamma)=\zeta_{n_\rho}^{n_\gamma}$. 
We need to check whether there exists an $N\in \N_\rho$ such that
$\zeta^{nN}=\zeta_{n_\rho}^{n_\gamma}$. Observe that, since $(n,n_\rho)=1$,
there is a unique solution $k$ to the congruence equation $n k =1 \mod n_\rho$. 
This is obtained by reducing modulo $n_\rho$ the relation $n k + n_\rho \ell =1$,
which is satisfied by a pair of $k,\ell \in \Z$, because $(n,n_\rho)=1$. Such $k$
is unique modulo $n_\rho$, since if $k'$ is another solution, $n (k-k')=0$ mod $n_\rho$
implies $n_\rho | (k-k')$ since $(n,n_\rho)=1$.
Note that $(k,n_\rho)$ divides $n k + n_\rho \ell$, hence $(k,n_\rho)=1$. 
Then $N=m k$ satisfies $\zeta^{nN}=\zeta^m$. 
For $(\gamma,\zeta) \in \hat\pi_\rho$, let $\delta_{(\gamma,\zeta)}$ be the corresponding
generator of the group ring $\Q[\hat\pi_\rho]$. The maps \eqref{sigmanhatpi} extend to
endomorphisms of $\Q[\hat\pi_\rho]$ by $\sigma_n(\delta_{(\gamma,\zeta)})=\delta_{(\gamma,
\zeta^n)}$. Since we clearly have $\sigma_n \circ \sigma_m =\sigma_{nm}$, the
maps \eqref{sigmanhatpi} determine a semigroup action of $\N_\rho$ by endomorphisms
of $\Q[\hat\pi_\rho]$. For the endomorphisms $\alpha_n: \Q[\hat\pi_\rho] \to \Q[\hat\pi_\rho]$
of \eqref{alphanhatpi} we also need to check that, for $(\gamma, \zeta)\in \hat\pi_\rho$ 
and $n\in \N_\rho$, if $\eta\in \mmu_\infty$ is such that $\eta^n =\zeta$, then $(\gamma,\eta)$
is also in $\hat\pi_\rho$. This can be seen immediately, since we know that there is some
$m\in \N_\rho$, such that $\zeta^m =\rho(\gamma)$, hence we also have 
$\eta^{nm}=\zeta^m=\rho(\gamma)$, hence $(\gamma,\eta)\in \hat\pi_\rho$. Thus,
the $\alpha_n$ of \eqref{alphanhatpi} are well defined. It is then also immediate to verify that we have
$$ \sigma_n \circ \alpha_n (\delta_{(\gamma,\zeta)}) =\frac{1}{n} \sum_{\eta\,:\, \eta^n =\zeta}  \sigma_n(\delta_{(\gamma,\eta)})= \frac{1}{n} \sum_{\eta\,:\, \eta^n =\zeta} \delta_{(\gamma,\zeta)}
= \delta_{(\gamma,\zeta)}, $$
$$ \alpha_n \circ \sigma_n (\delta_{(\gamma,\zeta)})=
\frac{1}{n} \sum_{\eta\,:\, \eta^n =\zeta^n} \delta_{(\gamma,\eta)} =
\frac{1}{n} \sum_{\xi\,:\, \xi^n =1} \delta_{(1,\xi)} \cdot \delta_{(\gamma,\zeta)}, $$
since solutions of $\eta^n=\zeta^n$ are of the form $\xi \zeta$. The element
$e_n =n^{-1} \sum_{\xi\,:\, \xi^n=1} \delta_{(1,\xi)}$ is an idempotent since we have
$$ e_n\cdot e_n = \frac{1}{n} \sum_{\xi_1\,:\, \xi_1^n=1} \frac{1}{n} \sum_{\xi_2\,:\, \xi_2^n=1} \delta_{(1,\xi_1 \xi_2)} =\frac{1}{n} \sum_{\chi\,:\, \chi^n=1} \delta_{(1,\chi)} = e_n. $$
\endproof

\smallskip

Thus, we can form the semigroup crossed product algebra as in the Bost--Connes case,
as a direct consequence of the previous proposition.

\begin{cor}\label{crossprodhatpi}
The semigroup crossed product algebra $\cA_{\hat\pi_\rho,\Q}:=\Q[\hat\pi_\rho]\rtimes_\alpha \N_\rho$ has
generators unitaries $\delta_{(\gamma,\zeta)}$, for $(\gamma,\zeta)\in \hat\pi_\rho$, and
isometries $\mu_n$, for $n\in \N_\rho$, satisfying 
$$ \mu_n^* \mu_n=1, \ \ \  \mu_n \mu_n^* =e_n, \ \ \ \mu_n \mu_m=\mu_{nm}, \ \ \ 
\mu_n \mu_m^* =\mu_m^* \mu_n \text{ for } (n,m)=1, $$ 
$$ \mu_n \delta_{(\gamma,\zeta)} \mu_n^* = \alpha_n (\delta_{(\gamma,\zeta)}), \ \ \ 
\mu_n^* \delta_{(\gamma,\zeta)} \mu_n =\sigma_n(\delta_{(\gamma,\zeta)}). $$
The $C^*$-algebra $C^*_r(\hat\pi_\rho)\times_\alpha \N_\rho$, with the same generators
and relations, is a $C^*$-algebra completion of $\cA_{\hat\pi_\rho,\Q}\otimes_\Q \C$.
\end{cor}

We can now combine this construction with the one described in the previous subsections
and define the $C^*$-algebra of observables of the combined system to be the following.

\begin{defn}\label{combinedCstar}
The $C^*$-algebra of observable of the quantum statistical mechanical
system of cyclic branched coverings of $S^3$ and knots is given by the
Bernoulli crossed product
\begin{equation}\label{combalg}
\bigotimes_{g\in \cG_\cK} \left( C^*_r(\hat\pi_\rho)\rtimes_\alpha \N_\rho \right) \rtimes \cG_\cK.
\end{equation}
\end{defn}

\begin{rem}\label{innercomb}{\rm
In the following we will refer to $C^*_r(\hat\pi_\rho)\rtimes_\alpha \N_\rho$ and its
associated quantum statistical mechanics as ``the inner system", and to \eqref{combalg}
as ``the combined system" or the ``total system". }
\end{rem}

\medskip
\subsection{Quantum statistical mechanics of the inner system}\label{QSMinnerSec}

By Lemma~\ref{projlimhatpiKn} and Definition~\ref{Nrho}, we have 
$\hat\pi_\rho \subset \pi\times \Q/\Z$ and $\N_\rho \subset \N$.  In
order to construct a quantum statistical mechanical system on 
$C^*_r(\hat\pi_\rho)\rtimes_\alpha \N_\rho$ that incorporates the
usual Bost--Connes dynamics, we start by considering the
algebra 
\begin{equation}\label{algBCpi}
C^*_r(\pi) \otimes C^*(\Q/\Z) \rtimes \N,
\end{equation} 
where $\N$ acts on $C^*(\Q/\Z)$ with the Bost--Connes endomorphisms 
\eqref{sigmanBC}, with $\cA_{BC}=C^*(\Q/\Z) \rtimes \N$ given in terms of
generators and relations as in \eqref{relBC1}, \eqref{relBC2}. On the algebra
\eqref{algBCpi}, we consider the time evolution $\sigma_t(\gamma\otimes a)=\gamma
\otimes \sigma_t(a)$, with $\gamma\in \pi$ and $a\in \cA_{BC}$, 
where $\sigma_t(a)$ is the Bost--Connes time evolution. 
We consider then representations of \eqref{algBCpi} on the Hilbert space
$\cH=L^2(\pi, \tau) \otimes \ell^2(\N)$, where $\tau$ is the von Neumann
trace on the group von Neumann algebra, with $\tau(1)=1$ and 
$\tau(\gamma)=0$, for $\gamma\neq 1$, given by
\begin{equation}\label{repBCpi}
\pi_u (\gamma \otimes a) \, \xi(\gamma')\otimes \epsilon_m = R(\gamma) \xi(\gamma') \otimes \pi_u(a) \epsilon_m, 
\end{equation}
for $\xi \in L^2(\pi, \tau)$ and $\epsilon_m$ the standard basis of $\ell^2(\N)$, 
where $R(\gamma)$ is the right regular representation of $C^*_r(\pi)$ on
$L^2(\pi, \tau)$ and $\pi_u(a)$ is the Bost--Connes representation \eqref{BCreps} of $\cA_{BC}$
on $\ell^2(\N)$.

\begin{lem}\label{repBCpiHam}
In the representation \eqref{repBCpi}, the time evolution $\sigma_t$ is implemented by
the Hamiltonian $H=1 \otimes H_{BC}$, where $H_{BC}$ is the Hamiltonian of the
Bost--Connes system. 
\end{lem}

\proof  For $H\, \xi(\gamma')\otimes \epsilon_m = \log(m)\, \xi(\gamma')\otimes \epsilon_m$,
we have $$e^{it H} \, \pi_u(\gamma \otimes a) \, e^{-it H} = R(\gamma) \otimes e^{it H_{BC}} \, \pi_u(a)\,
e^{-it H_{BC}} = R(\gamma) \otimes \pi_u(\sigma_t(a))= \pi_u(\sigma_t(\gamma \otimes a)).$$
\endproof

\begin{prop}\label{repBCpiKMS}
The functionals $\psi_\beta:=\tau \otimes \varphi_\beta$,
with $\tau$ the von Neumann trace and $\varphi_\beta$ a KMS$_\beta$ state of
the Bost--Connes system are KMS states of $(C^*_r(\pi) \otimes \cA_{BC}, \sigma)$.
Indeed, all KMS states are of this form.
\end{prop}

\proof  To see that the functionals $\psi_\beta =\tau \otimes \varphi_\beta$ satisfy the KMS$_\beta$
condition, consider elements $X,Y\in C^*_r(\pi)\otimes \cA_{BC}$ of the form 
$X=c \otimes a$ and $Y=c'\otimes a'$, with $c,c'\in C^*_r(\pi)$ and
$a,a'\in \cA_{BC}$. Then set $\tilde F_{X,Y}(z):=\tau(cc') F_{aa'}(z)$, where $F_{aa'}(z)$
is the holomorphic function expressing the KMS$_\beta$ condition for the state $\varphi_\beta$
on the algebra $\cA_{BC}$. The function $\tilde F_{X,Y}$ is clearly holomorphic on $\cI_\beta$
and continuous on $\partial\cI_\beta$ because $F_{aa'}(z)$ is. Moreover, it satisfies
$$\tilde F_{X,Y}(t)=\tau(cc') \varphi_\beta(a \sigma_t(a')) =\psi_\beta(X\sigma_t(Y))$$
$$ \tilde F_{X,Y}(t+i\beta)=\tau(cc') \varphi_\beta(\sigma_t(a')a) =\tau(c'c) \varphi_\beta(\sigma_t(a')a) 
=\psi_\beta(\sigma_t(Y)X),$$
hence it expresses the KMS$_\beta$ condition for $\psi_\beta$. Conversely, suppose
given a KMS$_\beta$ state $\psi_\beta$ for $(C^*_r(\pi) \otimes \cA_{BC}, \sigma)$.
It is known (see for instance \S 5.3.1 of \cite{BR}), that the KMS condition expressed
as above, in terms of interpolation of $\psi_\beta(X\sigma_t(Y))$ and $\psi_\beta(\sigma_t(Y)X)$
by a holomorphic function $F_{X,Y}(z)$, is equivalent to the property that, for all $X,Y$ in a dense
involutive subalgebra $\cA_{an}$ of ``analytic elements" (also called  ``entire elements")
the state satisfies $\psi_\beta(XY)=\psi_\beta(Y \sigma_{i\beta}(X))$. In particular,
for elements in  $\cA_{an}$ of the form $c\otimes 1$ and $c'\otimes 1$, we have
$\psi_\beta(cc' \otimes 1)=\psi_\beta(c'c \otimes 1)$. Indeed, since $\sigma_t(c\otimes 1)=c\otimes 1$
for $t\in \R$, elements of the form $c\otimes 1$ are always in $\cA_{an}$ with the analytic
extension of the time evolution still trivially given by $\sigma_z(c\otimes 1)=c\otimes 1$. Thus,
the KMS state $\psi_\beta$ restricted to elements of the form $c\otimes 1$ has to be a trace,
and therefore it has to agree with the unique von Neumann trace $\tau$. Consider then
elements of $\cA_{an}$ of the form $X=1 \otimes a$ and $Y=1\otimes b$, with $a,b\in \cA_{BC}$. 
Then $\sigma_t(X)=1\otimes \sigma_t(a)$ with $\sigma_t(a)$ the Bost--Connes time evolution.
Thus, the analytic continuation is also of the form $\sigma_z(X)=1\otimes \sigma_z(a)$,
that is, $a\in \cA_{an,BC}$ is an analytic element of the Bost--Connes algebra with
the corresponding analytic continuation of the time evolution. Thus, we have
$\psi_\beta(1\otimes ab)=\psi_\beta(1\otimes b\sigma_{i\beta}(a))$, which implies that, when 
restricted to $1\otimes \cA_{BC}$, the state satisfies $\psi_\beta(1\otimes a)=\varphi_\beta(a)$ 
for some KMS$_\beta$ state $\varphi_\beta$ of the Bost--Connes system.  Thus, for elements 
of the form $c\otimes a$, with $c\in C^*_r(\pi)$ and $a\in \cA_{BC}$ one obtains $\psi_\beta(c\otimes a)=\tau(c)\varphi_\beta(a)$. 
\endproof

\smallskip

\begin{rem}\label{KMShatpirho}{\rm 
When restricted to $C^*_r(\pi_\rho)\rtimes_\alpha \N_\rho$, the KMS$_\beta$ states
$\psi_\beta =\tau \otimes \varphi_\beta$ of $(C^*_r(\pi) \otimes \cA_{BC}, \sigma)$
define KMS states of the system $(C^*_r(\pi_\rho)\rtimes_\alpha \N_\rho,\sigma)$
with the induced time evolution.}
\end{rem}

\medskip
\subsection{Properties of the algebra of observables of the combined system}

We consider here the $C^*$-algebra \eqref{combalg} and the corresponding
von Neumann algebra
\begin{equation}\label{combvN}
\bigotimes_{g\in \cG_\cK} \left( \cN(\hat\pi_\rho)\rtimes_\alpha \N_\rho \right) \rtimes \cG_\cK,
\end{equation}
where $\cN(\hat\pi_\rho)$ is the group von Neumann algebra of $\hat\pi_\rho$. This
von Neumann algebra belongs to the class of noncommutative Bernoulli crossed products
\cite{Co74}. 

\smallskip

\subsubsection{Tensor product system}
In order to construct a quantum statistical mechanical system for this algebra,
compatible with the construction considered above for the inner system, we first
extend the construction of the inner system to the tensor product 
$$\otimes_{g\in \cG_\cK} C^*_r(\pi_\rho)\rtimes_\alpha \N_\rho.$$

\smallskip

Let $\cB_g = C^*_r(\pi_\rho)\rtimes_\alpha \N_\rho$ denote the $g$-th factor in
the above tensor product algebra.  An element $g\in \cG_\cK$ is an equivalence
class $g=K\ominus K'$ of pairs $(K,K')$ of knots, up to the equivalence defining
the Grothendieck group $\cG_\cK$ of the semigroup $(\cK,\#)$. 

\smallskip

On the algebra $\cB_{K\ominus K'}$ we consider a time evolution similar
to the one considered in \S \ref{QSMinnerSec}, induced on $C^*_r(\pi_\rho)\rtimes_\alpha \N_\rho$
from a time evolution $\sigma_t(\gamma \otimes a)=\gamma\otimes
\sigma_{t,K\ominus K'}(a)$ on $C^*_r(\pi)\otimes \cA_{BC}$, where for $a\in \cA_{BC}$ we now take
$\sigma_{t,K\ominus K'}(a)$ to be a scaled version of the Bost--Connes time evolution of the form 
$\sigma_{t,K\ominus K'} (e(r))=e(r)$
and 
\begin{equation}\label{BCscaledsigma}
\sigma_{t, K\ominus K'}(\mu_n) = n^{it \, f(K\ominus K')}\, \mu_n,
\end{equation}
where $f: \cG_\cK \to \R^*_+$ is a function, whose properties we specify below.
On the tensor product $\otimes_{g\in \cG_\cK} \cB_{g}$ we
consider the time evolution $\sigma_t =\otimes_{g} \sigma_{t,g}$.

\smallskip

\begin{defn}\label{Trtau}
For bounded linear operators on $L^2(\pi,\tau)\otimes \ell^2(\N_\rho)$ of the form
$R(\gamma) \otimes T$, we define $\Tr_\tau(R(\gamma) \otimes T):=\tau(\gamma) \Tr(T)$,
where $\Tr$ is the operator trace on $\cB(\ell^2(\N_\rho))$. The operator 
$R(\gamma) \otimes T$ is $\Tr_\tau$-class if $\Tr_\tau(R(\gamma) \otimes T)$ is finite,
that is, if $T$ is trace-class.
In particular, for a Hamiltonian of the form $1\otimes H$ as in Lemma \ref{repBCpiHam},
we define the partition function as
\begin{equation}\label{Ztau}
Z_\tau(\beta) = \Tr_\tau( 1\otimes e^{-\beta H} ).
\end{equation}
\end{defn}

\smallskip

\begin{prop}\label{ZetasigmaG}
Let $\cH =\otimes_{g\in \cG_\cK} \cH_g$ be the Hilbert space with 
$\cH_g = L^2(\pi,\tau)\otimes \ell^2(\N_\rho)$, with the algebra $\otimes_{g\in \cG_\cK} \cB_g$
acting on $\cH$ with the action $\pi_{u,g}$ of $\cB_g$ on $\cH_g$ as in \eqref{repBCpi}. Let
$H$ be the Hamiltonian implementing the time evolution $\sigma_t=\otimes_{g} \sigma_{t,g}$ 
in this representation. Consider a function $f: \cG_\cK \to \N$ with $f(g)=1$ for $g$ the class
of the unknot and $f(g)\geq 2$ for all other $g\in \cG_\cK$. Also assume that $f$ satisfies
\begin{equation}\label{fconvg}
\sum_{g\in \cG_\cK} f(g)^{-1} <\infty .
\end{equation}
Then the operator $e^{-\beta H}$ is $\Tr_\tau$-class if and only if $\beta>1$ 
and the partition function of the system is given by 
\begin{equation}\label{ZetaProdg}
Z_\tau (\beta) = \prod_{g\in \cG_\cK} \zeta_{n_\rho}(f(g)\beta) < \infty,
\end{equation}
where $\zeta_m(s)$ is the Riemann zeta function with the Euler
factors of primes $p$ with $p|m$ removed. 
\end{prop}

\proof For $\otimes_{g\in \cG_\cK} \cB_g$ represented on $\cH =\otimes_g \cH_g$
by the representation $\otimes_g \pi_{u,g}$, the time evolution $\sigma_t=\otimes_{g} \sigma_{t,g}$ 
is implemented on $\cH$ by a Hamiltonian of the form $H=\otimes_g (1\otimes H_g)$, 
where $H_g \epsilon_m =f(g)\, \log m \, \epsilon_m$, on the standard orthonormal basis 
$\{ \epsilon_m \}$ of $\ell^2(\N_\rho)$.
Definition~\ref{Trtau} extends to the case of a tensor product $\cH =\otimes_g \cH_g$ with
each $\cH_g=L^2(\pi,\tau)\otimes \ell^2(\N_\rho)$ and a Hamiltonian of the
form $H=\otimes_g (1\otimes H_g)$. For such an operator
we write in shorthand notation
\begin{equation}\label{expbetaHtens}
e^{-\beta H}=\otimes_g (1\otimes e^{-\beta H_g}).
\end{equation}
The trace is then given by
\begin{equation}\label{Trtautens}
\Tr_\tau(e^{-\beta H})=\prod_g \, \Tr_\tau( 1\otimes e^{-\beta H_g}) = \prod_g \Tr(e^{-\beta H_g}).
\end{equation}
On a given $\cH_g$ the Hamiltonian $H_g$ has 
$$ \Tr(e^{-\beta H_g}) = \sum_{n\in \N_\rho} n^{-f(g)\, \beta}, $$
which converges for $\beta > f(g)^{-1}$, since the sum is less than or equal to
$\sum_{n\geq 1} n^{-f(g)\, \beta}$, which converges for $\beta > f(g)^{-1}$ to
$\zeta(f(g)\beta)$ with $\zeta(s)$ the Riemann zeta function. Since the summation
is only on $\N_\rho$ instead of $\N$, the sum of the series is equal to
$$ \sum_{n\in \N_\rho} n^{-f(g)\, \beta} = \zeta_{n_\rho}(f(g)\beta), $$
where
$$ \zeta_m(s) =\prod_{p\not| m} (1-p^{-s})^{-1} = \sum_{n\in \N\,:\, (n,m)=1} n^{-s}. $$
Thus, the operator $1\otimes e^{-\beta H_g}$
is $\Tr_\tau$-class for $\beta > f(g)^{-1}$ and satisfies $\Tr_\tau(1\otimes e^{-\beta H_g})
=\zeta(f(g)\beta)$. Thus, the partition function of the system is given by the
infinite product
$$ \Tr_\tau(e^{-\beta H})=\prod_g \Tr_\tau(1\otimes e^{-\beta H_g}) =\prod_g \zeta_{n_\rho}(f(g)\beta). $$
The convergence of this depends on each of the factors $1\otimes e^{-\beta H_g}$ being
$\Tr_\tau$-class and on the convergence of the infinite product $\prod_g \zeta_{n_\rho}(f(g)\beta)$
in the range of $\beta$ where the $\Tr_\tau$-class condition is satisfied. Since 
$f(g)\geq 1$ for all $g\in \cG_\cK$, and each $1\otimes e^{-\beta H_g}$ is $\Tr_\tau$-class
for $\beta > f(g)^{-1}$, then all these operators are simultaneously $\Tr_\tau$-class in
the range $\beta >1$. In particular, since $\min_g f(g)=1$, then $\beta>1$ is exactly the range 
where $\Tr_\tau$-class condition holds. We then use the fact that the Riemann zeta
function satisfies
$$ \zeta(s) = \sum_{n\geq 1} n^{-s} = 1 + \sum_{n\geq 2} n^{-s} \leq 1 +
\int_2^\infty \frac{dx}{x} = 1 - \frac{1}{2(s-1)} \leq 1. $$
This gives, for $f(g)\geq 1$ and $\beta > 1$, 
$$ 0< \zeta_{n_\rho}(f(g)\beta)\leq \zeta(f(g)\beta) \leq 1 - \frac{1}{2(f(g)\beta -1)}. $$
Thus, the convergence of the infinite product $\prod_g \zeta(f(g)\beta)$ is
controlled by the convergence of the infinite product
$$ \prod_g \left( 1 - \frac{1}{2(f(g)\beta -1)} \right). $$
The convergence assumption \eqref{fconvg} implies the convergence
of $\sum_g (f(g)\beta-1)^{-1}$.
Recall that, for $a_\ell$ a sequence of complex numbers with $\sum_\ell |a_\ell|^2<\infty$
the convergence of the infinite product $\prod_\ell (1+a_\ell)$ is equivalent to the
convergence of the series $\sum_\ell a_\ell$. Since $f(g)\geq 2$ for all $g$ except the
unknot, for $\beta > 1$ we also have $(f(g)\beta-1)^{-1} <1$ for all $g$ except the
unknot. Thus, the convergence of $\sum_g (f(g)\beta-1)^{-1}$ also implies the
convergence of $\sum_g (f(g)\beta-1)^{-2}$. Thus the convergence of the series
$\sum_g (f(g)\beta-1)^{-1}$ is in fact equivalent to the convergence of the product
$\prod_g ( 1 - \frac{1}{2}(f(g)\beta -1)^{-1} )$. 
Thus, under the convergence assumption \eqref{fconvg}, we obtain that
the operator $e^{-\beta H}$ of \eqref{expbetaHtens} 
is $\Tr_\tau$-class in the range  $\beta>1$ and the partition function 
satisfies \eqref{ZetaProdg}.
\endproof

\smallskip

We have seen in Proposition~\ref{repBCpiKMS} how to obtain KMS$_\beta$
states $\psi_\beta =\tau\otimes \varphi_\beta$
on the algebra $\cB_g=C^*_r(\hat\pi_\rho)\rtimes_\alpha \N_\rho$ from
KMS states $\varphi_\beta$ of the Bost--Connes system and the von Neumann trace $\tau$. 
We focus now in particular on the extremal low temperature KMS states of the Bost--Connes system,
$\varphi_\beta =\varphi_{\beta,u}$ of \eqref{BClow}, with $u\in \hat\Z^*$. Let $\varphi_{\beta,u,g}$
denote an extremal low temperature KMS state for the Bost--Connes system with
Hamiltonian $H_g= f(g) H_{BC}$, where $H_{BC}$ is the restriction to $\ell^2(\N_\rho)$ of 
the usual Bost--Connes Hamiltonian, acting on $\ell^2(\N)$ by $H_{BC} \epsilon_m =\log(m) \epsilon_m$. 
Let $\psi_{\beta,u,g}=\tau\otimes \varphi_{\beta,u,g}$ be the corresponding KMS$_\beta$ state
on the system $(\cB_g, \sigma_{t,g})$.

Given a function $F: \cG_\cK \to \cB(\cH)$, for a fixed Hilbert space $\cH$, consider
the operator $\otimes_{g\in \cG_\cK} F(g)$ acting on the product
$\otimes_{g\in \cG_\cK}\cH_g$ with $\cH_g=\cH$, for all $g\in \cG_\cK$. 
In particular, we can write elements in the algebra $\otimes_{g\in \cG_\cK} \cB_g$
in the form of functions $$F: \cG_\cK \to \cB(L^2(\pi,\tau)\otimes \ell^2(\N_\rho)).$$
Then Proposition~\ref{ZetasigmaG} implies that we obtain a KMS$_\beta$ state on 
the tensor product system $(\otimes_{g\in \cG_\cK} \cB_g, \sigma_t=\otimes_g \sigma_{t,g})$
as follows.

\begin{cor}\label{tensorKMSf}
Let $f: \cG_\cK \to \N$ be a function satisfying the same hypotheses as in Proposition~\ref{ZetasigmaG}. 
Then $\Psi_{\beta,u,f}=\otimes_g \psi_{\beta,u,g}$ is a KMS$_\beta$ state on the crossed product
system $(\otimes_{g\in \cG_\cK} \cB_g, \sigma_t=\otimes_g \sigma_{t,g})$. It is explicitly given in the
Gibbs form
\begin{equation}\label{PsiGibbs}
\Psi_{\beta,u,f}(F) = \frac{\Tr_\tau( e^{-\beta \, f\, H_{BC}} F )}{Z_\tau(\beta)},
\end{equation}
with $Z_\tau(\beta)$ as in \eqref{ZetaProdg}. 
\end{cor} 

\proof As in Proposition~\ref{ZetasigmaG}, we have 
$$ \Tr_\tau( e^{-\beta \, f\, H_{BC}} F )=\prod_g \Tr_\tau((1\otimes e^{-\beta H_g}) F(g)). $$
Moreover, for an element $F(g)=1\otimes a_g$, with $a_g\in \cA_{BC}$ represented via the 
representation $\pi_u$, the above is equal to $\prod_g \Tr(e^{-\beta H_g} \pi_u(a_g))$ and one obtains
$$ \prod_g \frac{\Tr(e^{-\beta H_g} \pi_u(a_g))}{\zeta(f(g)\beta)} = \prod_g \varphi_{\beta,u,g}(a_g). $$
\endproof

\smallskip

\subsubsection{Bernoulli crossed product}
As above, given a function $F: \cG_\cK \to \cB(\cH)$, we consider
the operator $\otimes_{g\in \cG_\cK} F(g)$ on 
$\otimes_{g\in \cG_\cK}\cH_g$, with $\cH_g=\cH$, for all $g\in \cG_\cK$. Consider the
action of the group $\cG_\cK$ on the set of functions $F: \cG_\cK \to \cB(\cH)$ given by
\begin{equation}\label{alphagF}
\alpha_h(F)(g):=F(h^{-1}g), \ \ \ \text{ for } \ \ h,g\in \cG_{\cK}.
\end{equation} 
As above, we write elements in the algebra $\otimes_{g\in \cG_\cK} \cB_g$ in this way.

\smallskip

\begin{prop}\label{tevolcrossprod}
The time evolution determined by \eqref{BCscaledsigma} on the algebra $\otimes_g \cB_g$
extends to the crossed product algebra $(\otimes_g \cB_g) \rtimes \cG_\cK$ by setting
\begin{equation}\label{Uhtevolf} 
\sigma_t(U_h)= e^{it (f- \alpha_h(f)) \, H_{BC}} U_h,
\end{equation}
where $U_h$, for $h\in \cG_\cK$ are the unitaries implementing the crossed
product action $\alpha_h(F)=U_h F U_h^*$ for $F=\otimes_g F(g)\in \otimes_g \cB_g$.
\end{prop}

\proof
Let $\H:  \cG_\cK \to \cB(\cH)$ be the function $\bH(g)=1\otimes H_g = 1\otimes f(g) \, H_{BC}$,
with $H_{BC}\in \cB(\ell^2(\N_\rho))$ the Bost--Connes Hamiltonian. 
We then write the time evolution on functions $F: \cG_\cK \to \cB(\cH)$ as
\begin{equation}\label{tevolf}
\sigma_t (F) (g) = e^{it\H(g)} F(g) e^{-it \H(g)}.
\end{equation}
For $h\in \cG_\cK$, let $U_h$ be the unitary operator on 
$\otimes_{g\in \cG_\cK}\cH_g$, with $\cH_g=L^2(\pi,\tau)\otimes \ell^2(\N_\rho)$, which
acts as $(U_h \xi)_g =\xi_{hg}$, 
where we write elements of  $\otimes_{g\in \cG_\cK}\cH_g$ as $\xi=\otimes_g \xi_g$,
with $\xi_g\in L^2(\pi,\tau)\otimes \ell^2(\N_\rho)$. We then have
$U_h F U_h^* \xi = \alpha_h(F) \xi$.
This action satisfies
$$ U_h \sigma_t ( F ) U_h^* =U_h  e^{it\H} F e^{-it \H} U_h^*=
\alpha_h(e^{it\H} F e^{-it \H}), $$
where $(\alpha_h(e^{it\H} F e^{-it \H}) \xi)_g = e^{it f(h^{-1}g) H_{BC}} 
F(h^{-1}g) e^{-it f(h^{-1}g) H_{BC}} \xi_g$. On the other hand, we have
$$ \sigma_t (  U_h F  U_h^*) =  e^{it\H} U_h F U_h^* e^{-it \H} = e^{it\H} \alpha_h(F) e^{-it \H}, $$
where $(e^{it\H} \alpha_h(F) e^{-it \H} \xi)_g =  e^{it f(g) H_{BC}} 
F(h^{-1}g) e^{-it f(g) H_{BC}} \xi_g$. This implies that the action of $\cG_\cK$ transforms
the time evolution as $\sigma_{h,t}:=\alpha_h(\sigma_t)$ with
\begin{equation}\label{htevolf}
\sigma_{h,t}(F)(g) = e^{it\alpha_h(\H)(g)} F(g) e^{-it \alpha_h(\H)(g)}.
\end{equation}
Moreover, we obtain \eqref{Uhtevolf}, since
$$ \sigma_t(U_h)= e^{it\H} U_h e^{-it\H} = e^{it\H} e^{-it \alpha_h(\H)} U_h =
e^{itf \, H_{BC}} e^{-it \alpha_h(f)\, H_{BC}} U_h. $$
This determines how the time evolution extends to the crossed product 
$(\otimes_g \cB_g) \rtimes \cG_\cK$.
\endproof

\smallskip

Let $\psi_{\beta,g}$ denote a KMS$_\beta$ state, obtained as in Remark~\ref{KMShatpirho}, 
for the system $(\cB_g,\sigma_{t,g})$,
where $\sigma_{t,g}$ is the time evolution \eqref{BCscaledsigma} with Hamiltonian
$H(g)=f(g) \, H_{BC}$, and the algebra is $\cB_g=C^*_r(\hat\pi_\rho)\rtimes_\alpha \N_\rho$
as above. We denote by $\Psi_{\beta,u,f}$ the KMS$_\beta$ state on the system
$(\otimes_g \cB_g, \otimes_g \sigma_{t,g})$ determined by the $\psi_{\beta,u,g}$ as
in Corollary~\ref{tensorKMSf}.

\begin{lem}\label{KMSalphah}
Under the action $\alpha_h$ of $h\in \cG_\cK$, the KMS$_\beta$ state
$\Psi_{\beta,u,f}$ of Corollary~\ref{tensorKMSf} satisfies
\begin{equation}\label{alphahKMS}
\Psi_{\beta,u,f} \circ \alpha_h = \Psi_{\beta, u,\alpha_{h^{-1}}(f)}. 
\end{equation}
\end{lem}

\proof We have
$$  \Psi_{\beta,u,f}(\alpha_h(F))= \Psi_{\beta,u,f} (U_h F U_h^*) = \Psi_{\beta,u,f} ( \sigma_{-i\beta}(U_h^*) U_h F )
= \Psi_{\beta,u,f} ( e^{-\beta (\alpha_{h^{-1}}(f) -f ) H_{BC}} F ). $$
On the other hand, we also have
$$ \Psi_{\beta, u,\alpha_{h^{-1}}(f)} (F) 
= \frac{\Tr_\tau(e^{-\beta \alpha_{h^{-1}}(f) H_{BC}} F)}{Z_\tau(\beta)} $$
$$ = \frac{\Tr_\tau(e^{-\beta \, f\, H_{BC}} e^{-\beta (\alpha_{h^{-1}}(f) -f) H_{BC}} F)}{Z_\tau(\beta)} =
\Psi_{\beta,u,f}(e^{-\beta (\alpha_{h^{-1}}(f) -f ) H_{BC}} F ). $$
\endproof

\medskip
\subsection{Knot invariants and the function $f(g)$}\label{fgSec}
We now show how to construct a function $f: \cG_\cK \to \N$ that
satisfies the hypotheses of Proposition~\ref{ZetasigmaG} and
Corollary~\ref{tensorKMSf}, using knot invariants. 
As in \S \ref{statKaSec}, we write elements of $\cG_\cK$ in terms of
primary decomposition. 
Let $K\ominus K'=(a_1 K_1\#\cdots \# a_j K_j)\ominus 
(b_1 K_1'\# \cdots \# b_\ell K'_\ell)$ be an element of $\cG_\cK$ with primary
decompositions $K=a_1 K_1\#\cdots \# a_m K_m$ and $K'=b_1 K_1'\# \cdots \# b_\ell K'_\ell$,
where the $K_i$ and $K'_j$ are all distinct prime knots, with multiplicities $a_i$ and $b_j$.
Since we eliminate all possible common factors from the primary decomposition of
$K$ and $K'$, this description of elements $g=K\ominus K'\in \cG_\cK$ is unique.
We also use, as in \S \ref{statKaSec}, the notation $\omega(K)$ for 
the number of distinct prime knots in its primary decomposition of a knot $K$. 
It is then convenient to consider knot invariants that are additive under
connected sums, and for which there is a good estimate of the rate of
growth of the multiplicities. 

\smallskip

To this purpose, we proceed as in \S \ref{statKaSec}, and we restrict from the
Grothendieck group $\cG_\cK$ of the semigroup $(\cK,\#)$ of all knots with
the connected sum operation, to the subsemigroup $(\cK_a,\#)$ of alternating
knots and its Grothendieck group $\cG_{\cK,a}$, so that we can again use the
genus and the crossing numbers as invariants. This means that, for the purpose
of this section, we will be restricting to the Bernoulli crossed product
\begin{equation}\label{alterBern}
(\otimes_{g\in \cG_{\cK,a}} \cB_g) \rtimes \cG_{\cK,a},
\end{equation}
where, as before, $\cB_g = C^*_r(\hat\pi_\rho)\rtimes_\alpha \N_\rho$.

\smallskip

\begin{prop}\label{fgprop}
For $K\ominus K' \in \cG_{\cK,a}$, represented through its primary decomposition
$K\ominus K'=(a_1 K_1\#\cdots \# a_j K_j)\ominus 
(b_1 K_1'\# \cdots \# b_\ell K'_\ell)$ with no common prime factors, the
function 
\begin{equation}\label{fgeq}
f(K\ominus K')= q^{\lceil \beta_+ \rceil\left(\sum_{i=1}^m a_i (Cr(K_i)+g(K_i)) + \sum_{j=1}^\ell 
b_j (Cr(K'_j)+g(K'_j)) \right)},
\end{equation}
with $\lceil \beta_+ \rceil$ the smallest integer greater than or equal to the value 
$\beta_+$ of Theorem~\ref{ZetaCrg},
satisfies the hypotheses of Proposition~\ref{ZetasigmaG} and
Corollary~\ref{tensorKMSf}.
\end{prop}

\proof The function $f(g)$ takes values in $\N$, since $q\geq 2$ is a fixed
integer, and it takes value $f(g)=1$ only when $g$ is the unknot, since only in that
case the exponent is zero. Thus, we only need to check that the convergence
property $\sum_g f(g)^{-1} <\infty$ is satisfied. By Theorem~\ref{ZetaCrg} we
know that
$$ \sum_{K \in \cK_a} f(K)^{-1} <\infty, $$
where $f(K)=q^{\lceil \beta_+ \rceil ( Cr(K) +g(K))}$,
while by \eqref{ZetaHGKa} and Proposition~\ref{propHGKa} we see that also
$$ \sum_{K\ominus K'\in \cG_{\cK,a}} f(K\ominus K')^{-1} < \infty. $$
\endproof

\smallskip

In particular, we can then see more explicitly the action
$f\mapsto \alpha_{h^{-1}}(f)$ that determines the transformation
property of the KMS$_\beta$ state $\Psi_{\beta,u,f}$ as in 
Lemma~\ref{KMSalphah}. 

\smallskip

\begin{cor}\label{alphafcor}
For $h=\pm K$ in $\cP_a$, the action $f(g)\mapsto \alpha_{h^{-1}}(f)(g)$
raises or lowers by one the multiplicity of the prime factor $K$
in the primary decomposition of $g=K\ominus K'=(a_1 K_1\#\cdots \# a_j K_j)\ominus 
(b_1 K_1'\# \cdots \# b_\ell K'_\ell)$.
\end{cor} 

\proof It suffices to see the effect of the
action of an element $h\in \cG_{\cK,a}$ 
given by a single prime knot $K \in \cP_a$
with either a positive or a negative exponent. 
This gives either
$$ \alpha_{K}(f) (K_1\ominus K_2)= f(K_1\# K \ominus K_2), $$
or, respectively, 
$$ \alpha_{-K}(f) (K_1\ominus K_2)= f(K_1 \ominus K_2\# K). $$
Since the definition of the function $f$ depends on the primary
decomposition of $K_1\ominus K_2$ without common factors,
the result depends on whether $K$ is a prime factor of either $K_1$
or $K_2$. By analogy to the case of integers, for a knot $K$, we denote by
$(K_i,K)$ the connected sum of all the prime factors (with multiplicity)
common to $K_i$ and $K$ and we denote by $K_i/(K_i,K)$ the result
of removing $(K_i,K)$ from the primary decomposition of $K_i$. Since
$K$ is a single prime knot, $(K_i,K)=K$ if it is non-trivial, that is, if
$K|K_i$ and it is the trivial knot otherwise. Similarly $K_i/(K_i,K)=K_i/K$
in the first case and $K_i/(K_i,K)=K_i$ in the second. Note that,
if $K_1\ominus K_2$ is represented in a primary decomposition without
common factors, then $K$ can divide either $K_1$ or $K_2$ or neither,
but it cannot divide both. Thus, the result of 
$\alpha_{\pm K}(f) (K_1\ominus K_2)$ is simply to lower or rise by one the
power of $K$ in the primary decomposition.
\endproof

\smallskip

\begin{rem}\label{QuesAlex}{\rm It would be interesting to see if the
construction presented in this paper can be extended to incorporate
other, more sophisticated invariants of knots. For example, the type of
(twisted) $L^2$-Alexander-Conway invariants of knots considered
in \cite{LiZhang}, \cite{LiZhang2} are naturally defined in terms
of the von Neumann algebra $\cN(\pi_K)$ of the knot group and
appear to be suitable for the quantum statistical mechanical
setting considered here.}
\end{rem}

\bigskip

\subsection*{Acknowledgement} The first author is supported by NSF grants 
DMS-1201512 and PHY-1205440. The second author is supported by a
Summer Undergraduate Research Fellowship at Caltech.

\end{document}